\documentclass[arps,prd,onecolumn,showpacs]{revtex4}
\usepackage{graphicx}
\usepackage{amsmath}
\usepackage{amsfonts}
\usepackage{amssymb}
\usepackage{color}
\usepackage{epsf}

\newcommand{\be}{\begin{equation}}         
\newcommand{\ee}{\end{equation}}
\newcommand{\ba}{\begin{eqnarray}}
\newcommand{\ea}{\end{eqnarray}}
\newcommand{\nn}{\nonumber}

\newcommand\lsim{\mathrel{\rlap{\lower4pt\hbox{\hskip1pt$\sim$}}
        \raise1pt\hbox{$<$}}}
\newcommand\gsim{\mathrel{\rlap{\lower4pt\hbox{\hskip1pt$\sim$}}
        \raise1pt\hbox{$>$}}}
\def\k{{\bf k}}
\def\x{{\bf x}}

\begin{document}

\title{Spherical collapse in $\nu \Lambda CDM$}
\author{ Marilena LoVerde}
\affiliation{Enrico Fermi Institute, Kavli Institute for Cosmological Physics, Department of Astronomy and Astrophysics, University of Chicago, Illinois, 60637, USA }
\begin{abstract}
The abundance of massive dark matter halos hosting galaxy clusters provides an important test of the masses of relic neutrino species. The dominant effect of neutrino mass is to lower the typical amplitude of density perturbations that eventually form halos, but for neutrino masses $\gsim 0.4eV$ the threshold for halo formation can be changed significantly as well. We study the spherical collapse model for halo formation in cosmologies with neutrino masses in the range $m_{\nu i } =0.05eV$- $1eV$ and find that halo formation is differently sensitive to $\Omega_\nu$ and $m_\nu$. That is,  different neutrino hierarchies with common $\Omega_\nu$ are in principle distinguishable. The added sensitivity to $m_{\nu }$ is small but potentially important for scenarios with heavier sterile neutrinos. Massive neutrinos cause the evolution of density perturbations to be scale-dependent at high redshift which complicates the usual mapping between the collapse threshold and halo abundance. We propose one way of handling this and compute the correction to the halo mass function within this framework. For $\sum m_{\nu i} \lsim 0.3eV$, our prescription for the halo abundance is only $\lsim 15\%$ different than the standard calculation. However for larger neutrino masses the differences approach $50-100 \%$ which, if verified by simulations, could alter neutrino mass constraints from cluster abundance. 
\end{abstract}
\maketitle

\section{Introduction}
\label{sec:intro}
The exquisite measurements of temperature anisotropies in the cosmic microwave background (CMB) by Planck \cite{Ade:2013zuv}, WMAP, \cite{Hinshaw:2012aka}, SPT \cite{ Hou:2012xq} and ACT \cite{Sievers:2013ica} reveal a universe that, on large scales, can be remarkably well characterized by just a few cosmological parameters. One parameter that is not currently required is the energy density in massive relic neutrinos. Cosmological evidence for massive neutrinos is, at present undetected, and as such is considered an extension to the $\Lambda CDM$ model. Neutrino oscillation experiments require that massive neutrinos contribute at least a few tenths of a percent to the cosmic energy budget today, $\Omega_\nu h^2 \gsim 0.06eV/94eV$ \cite{Beringer:1900zz}. 

The presence of massive neutrinos changes the evolution of matter perturbations. Matter perturbations with wavelengths smaller than the neutrino free-streaming length are suppressed and a detection of this suppression would provide a measure of the energy density in relic neutrinos (for a review see \cite{Lesgourgues:2006nd}). The neutrino-induced suppression to the matter power spectrum scales primarily with $\sum_i m_{\nu i}$ and current bounds on the sum of the neutrino masses from cosmological datasets are $\sum_i m_{\nu i} \lsim 0.2 eV - 1eV$ (see e.g. \cite{Ade:2013zuv, Hinshaw:2012aka, Hou:2012xq, Sievers:2013ica} for CMB constraints, \cite{Seljak:2006bg, Reid:2009nq, Thomas:2009ae, Saito2010, Swanson:2010sk, Xia:2012na, RiemerSorensen:2011fe,Zhao:2012xw,dePutter:2012sh, Beutler:2013yhm} for constraints from galaxy and Lyman-alpha forest surveys, and \cite{Vikhlinin:2008ym,Mantz:2009rj,Benson:2011uta,Reichardt:2012yj,Hasselfield:2013wf,Ade:2013lmv} for constraints from the abundance of galaxy clusters). 

The standard model of particle physics includes three active neutrino species and three neutrino mass eigenstates. However, there are a number of anomalies in particle physics, nuclear physics, and astrophysics datasets that suggest the presence of additional light neutrino species with mass $\sim 1eV$ (for a review see \cite{Abazajian:2012ys,Conrad:2013mka} and also \cite{Wyman:2013lza,Hamann:2013iba,Battye:2013xqa}). The invisible decay width of the $Z$ boson limits the number of weakly interacting neutrinos to three, so if an additional neutrino species exists it must be sterile \cite{Beringer:1900zz}. Cosmological datasets bound the effective number of all (active and sterile) neutrinos species (i.e. the number of relativistic fermionic degrees of freedom in the early universe). Current constraints from Planck are $2.72 < N_{eff} < 4.04$ at $95\%$ confidence \cite{Ade:2013zuv}.

 As discussed in \cite{LoVerde:2013lta} neutrinos with masses $m_{\nu } \gsim 0.2eV$ (and the standard temperature of $T_\nu\sim 1.7\times 10^{-4}eV$) cluster much more strongly around dark matter halos so one might expect correspondingly larger effects on halo formation and abundance. While the evidence for an additional, more massive, sterile species is far from strong, one motivation for this work is to provide a framework for understanding the signatures of neutrinos with masses as large as $1eV$ on halo formation and abundance.   In any case, massive neutrinos exist and are one of two known components of dark matter in the universe today. It is therefore important to understand how calculations of cold dark matter (CDM) structure formation are altered by the neutrino component. 

In this paper, we consider the effects of massive neutrinos on the simplest, spherical collapse model of halo formation \cite{Gunn:1972sv}. The effects of massive neutrinos on spherical collapse were first studied by Ichiki and Takada \cite{Ichiki:2011ue} and this paper largely follows their approach. The  main differences between this work and \cite{Ichiki:2011ue} are that (i) we start the spherical collapse calculations at later times when the perturbations are subhorizon using the initial conditions from the publicly available CAMB code \cite{Lewis:1999bs} and (ii) we allow for multiple massive neutrino species and larger individual neutrino masses than those considered in \cite{Ichiki:2011ue}. Including multiple massive species allows us to separately study the effects of $m_\nu$ and $\Omega_\nu$ on spherical collapse. Larger neutrino masses also cause greater changes to the evolution of spherical overdensities which is helpful to understand the different physical effects and ultimately may be important to model and constrain a sterile species. 

The semi-analytic spherical collapse model we consider here is, of course, not a precise description of cold dark matter structure formation in the universe. The current community standard for modeling structure formation is high-resolution N-body simulations (see e.g. \cite{Tinker:2008ff} and references therein). Nevertheless, the spherical collapse model is a useful testing ground for developing an understanding of how non-standard cosmologies impact halo formation (e.g. \cite{Lahav:1991wc,Barrow:1993,Wang:1998gt, Basse:2010qp, Creminelli:2009mu,Schmidt:2009yj,LoVerde:2011iz}). Furthermore, at present there are only a handful of N-body simulations that include the effects of massive neutrinos \cite{Colin:2007bk,Brandbyge:2008rv,Agarwal:2010mt, Viel:2010bn,Viel:2010bn,Brandbyge:2010ge,Marulli:2011he,VillaescusaNavarro:2012ag,Upadhye:2013ndm} and even fewer that include both cold dark matter and neutrino particles in simulations \cite{Colin:2007bk,Brandbyge:2008rv,Viel:2010bn,Brandbyge:2010ge,VillaescusaNavarro:2012ag, Villaescusa-Navarro:2013pva, Castorina:2013wga,Costanzi:2013bha}. In the mixed-dark-matter simulations of \cite{Brandbyge:2010ge, Marulli:2011he, VillaescusaNavarro:2012ag} the Sheth-Tormen model for the halo mass function \cite{Sheth:1999mn} was found to continue to describe the halo abundance in a cosmology with massive neutrinos with $m_{\nu i } \lsim 0.2 eV$ provided one makes the replacement  $\Omega_m\rightarrow  \Omega_c + \Omega_b$ (i.e. the neutrino contribution to $\Omega_m$ is neglected). In \cite{Costanzi:2013bha} the same replacement was found to give good agreement between the $\nu CDM$ mass functions and the Tinker spherical-overdensity mass function \cite{Tinker:2008ff} as long as the CDM power spectrum was used to calculate the variance of mass fluctuations on scale $M$ (rather than the total matter power spectrum). These descriptions of the change to the halo mass function are in qualitative agreement with our calculations that show that the collapse threshold is changed by $\lsim 0.5\%$  for $\sum_i m_{\nu i }\lsim 0.3 eV$ and $M \le 10^{15} M_\odot$ (see also \cite{Ichiki:2011ue}). It would be interesting to compare our calculations which include larger values of neutrino masses to mixed dark matter N-body simulations. 

In the plots and numerical examples shown throughout this paper we use Planck \cite{Ade:2013zuv} values of the standard, flat $\Lambda CDM$ cosmological parameters: Hubble parameter $h = 0.67$, cold dark matter (CDM) density  $\Omega_c h^2 = 0.1199 $ and baryon density $\Omega_bh^2 =  0.022 $. When solving for the evolution of spherical overdensities we treat baryons and CDM as a single fluid (for a discussion of baryon effects on spherical collapse calculations see \cite{Naoz:2006ye, Ichiki:2011ue}). We assume three species of massive neutrinos with variable masses $m_{\nu 1}$, $m_{\nu 2}$ and $m_{\nu 3}$. Massive neutrinos contribute a fraction $\Omega_{\nu }h^2 \approx \sum m_{\nu i}/(94 eV)$ to the critical energy density so for fixed CDM and baryon densities, changing the neutrino masses leads to a different total matter ($\Omega_m = \Omega_c + \Omega_b +  \Omega_\nu$) densities today. We adjust $\Omega_\Lambda$ to keep the universe flat, that is $\Omega_\Lambda = 1 - \Omega_c - \Omega_b - \Omega_\nu - \Omega_\gamma$ and these plots are referred to as at fixed $\Omega_c$.  In this paper, ``Inverted Hierarchy" means $m_{\nu 1} = m_{\nu 2} = 0.05eV$ and $m_{\nu} = 0eV$, and ``no massive $\nu$" means $m_{\nu 1} = m_{\nu 2} = m_{\nu 3} = 0eV$. We also make comparisons between cosmologies with massive neutrinos and a cosmology with massless neutrinos and the same total matter density, $\Omega_c^{comparison} = \Omega_c + \Omega_\nu$, we refer to the difference between these two cases as changes with fixed $\Omega_m$. In numerical examples, we consider several representative scenarios for the neutrino mass hierarchy which are not all compatible with oscillation data, but they allow us to compare scenarios with a fixed $\Omega_\nu$ but different individual neutrino masses. We solve for the background cosmology and spherical halo collapse self-consistently for each set of neutrino masses. 

The rest of this paper is organized as follows. In \S\ref{sec:nuLCDM} we outline our method for solving spherical collapse in $\nu\Lambda CDM$ and present numerical calculations of the halo evolution in this model.  In \S \ref{sec:massfcn} we relate spherical collapse results to halo abundance in $\nu\Lambda CDM$ cosmologies. Massive neutrinos cause the evolution of density perturbations to depend on scale and this fact leads to some subtleties in relating spherical collapse calculations to halo abundance so we review the framework in detail.   Numerical results for the neutrino-induced changes to the collapse threshold and halo abundance are presented in the same section. Conclusions are given in  \S \ref{sec:conclusions}.  Appendix \ref{sec:ICsfromCAMB} describes how we set up the initial conditions using the output of CAMB \cite{Lewis:1999bs}. Appendix \ref{sec:deltaMnutests} presents tests of whether our results are sensitive to the approximation used to calculate the neutrino mass that clusters inside the halo during collapse (they are not). Appendix \ref{sec:comparedeltacrit} further explores the meaning of the collapse threshold in $\nu \Lambda CDM$ and presents a comparison of several approaches to defining this quantity.

\section{Spherical collapse in a $\nu  \Lambda CDM$ universe}
\label{sec:nuLCDM}
In this section we develop the spherical collapse model for halo formation in a universe with multiple components to the energy density. Our halo is a homogenous spherical overdensity in CDM and baryons that accumulates neutrino mass as the amplitude of the CDM and baryon perturbation grows during gravitational collapse. 

The unperturbed background universe evolves according to the Friedmann equation, 
\be
\label{eq:Friedmann}
H^2(a)=\frac{8\pi G}{3}\left(\bar{\rho}_c(a)+\bar{\rho}_b(a)+\bar{\rho}_\nu(a)+\bar{\rho}_\gamma(a)+\bar{\rho}_\Lambda\right)
\ee 
where $a$ is the scale factor and we have included CDM $\bar\rho_c$, baryons $\bar\rho_b$, neutrinos $\bar\rho_\nu$, photons $\bar\rho_\gamma$, and a cosmological constant $\bar\rho_\Lambda$. The number density, energy density, and pressure of a single neutrino mass eigenstate with mass $m_{\nu i}$ is given by
\be
\label{eq:nbarnu}
\bar{n}_{\nu i} = 2 \int \frac{d^3 p}{(2\pi)^3}\frac{1}{e^{p/T_\nu}+1}\,,\quad\bar{\rho}_{\nu i} = 2 \int \frac{d^3 p}{(2\pi)^3}\frac{\sqrt{p^2 + m_{\nu i}^2}}{e^{p/T_\nu}+1}\,,\quad\bar{P}_{\nu i} = 2  \int \frac{d^3 p}{(2\pi)^3}\frac{p^2}{3\sqrt{p^2 + m_{\nu i}^2}}\frac{1}{e^{p/T_\nu}+1}\,.
\ee
where the neutrino temperature is $T_\nu(a)  = 1.95491 K/a$ and the total neutrino energy density is $\bar\rho_\nu = \sum_i \bar{\rho}_{\nu i}$. The other components evolve as $\bar{\rho}_c\propto 1/a^3$, $\bar{\rho}_b\propto 1/a^3$, $\bar{\rho}_\gamma \propto 1/a^4$, and $\bar{\rho}_\Lambda = const.$. 
\subsection{Equation of motion for $R$}
\label{ssec:Reom}
We follow the evolution of the radius $R(t)$ enclosing a constant CDM and baryon mass $M$. We start our calculations at $z_{init} = 200$.  At this redshift the density perturbations that form halos at late times have amplitudes $\delta_{init} \sim \mathcal{O}(10^{-2})$ so we may set the initial conditions using linear theory (see Appendix \ref{sec:ICsfromCAMB} for further discussion).  However, $z_{init} = 200$ is late enough that the perturbations of interest are all well within the horizon. Furthermore, perturbations in the baryon density have nearly caught up with perturbations of the CDM so we will treat them as a single fluid with energy density $\bar{\rho}_{cb} \equiv \bar{\rho}_c + \bar{\rho}_b$. The CDM and baryon density perturbations are given by 
\be
\delta_{cb} \equiv \frac{\bar\rho_c \delta_c + \bar\rho_b\delta_b}{\bar\rho_c +\bar\rho_b}\,.
\ee
 The equation of motion for a mass shell of radius $R$ enclosing constant (CDM + baryon) mass $M$ is 
\be
\label{eq:ddotRall}
\ddot{R}=-\frac{GM}{R^2}-\frac{4\pi G \int_0^{R} dr r^2 (\rho_{rest}(r,t)+3P_{rest}(r,t))}{R^2}\,,
\ee
where $\dot{}$ indicates a derivative with respect to time, $\rho_{rest}$ and $P_{rest}$ are the energy density and pressure of photons, neutrinos, and cosmological constant. 
For a halo of mass $M$, we use the linear initial conditions for $R$ 

\be
\label{eq:Rics}
R_{init} = \bar{R}_{init}\left(1 - \frac{1}{3}\delta_{cb, init}\right) \,, \qquad \dot{R}_{init} = H_{init}\bar{R}_{init}\left(1 - \frac{1}{3}\delta_{cb, init} - \frac{1}{3}H_{init}^{-1}\dot{\delta}_{cb, init}\right) \,,\qquad \bar{R}_{init} =  \left(\frac{3 M}{4\pi\bar\rho_{cb}}\right)^{1/3}\,
\ee
where $\,\bar{\,}\,$ indicates unperturbed quantities and we use the subscript $init$ to indicate quantities at the initial redshift $z_{init}$. In a cosmology with massive neutrinos, the growth of linear density perturbations is different for perturbations with wavelengths above and below the neutrino free-streaming scale. In this paper we will use CAMB to find $\dot{\delta}_{cb}/\delta_{cb}$ at the initial time but it is instructive to study the analytic expressions for an $\Omega_m=1$ universe to get a sense of how massive neutrinos affect the initial conditions for $R(t)$.  In an exactly matter-dominated phase the linear growing mode with wavenumber $k$ evolves as
\be
\delta_{cb}(k, a) \propto \left\{\begin{array}{ccc} a& {\rm for} & k \lsim k_{{\rm {\tiny free-streaming}}} \\
a^{1-3/5f_\nu}& {\rm for} & k \gsim k_{{\rm {\tiny free-streaming}}} \end{array}\right.
\ee
where $k_{{\rm {\tiny free-streaming}}}\sim aH(a)m_\nu/T_{\nu}(a)$ and $f_\nu \equiv \Omega_\nu/(\Omega_c+\Omega_b+\Omega_\nu)$. The perturbations that form halos at late times are predominantly on scales smaller than the neutrino free-streaming scale. Therefore we expect that the presence of massive neutrinos will delay halo formation. 

The evolution of $R(t)$ shown in Eq.~(\ref{eq:ddotRall}) is sensitive to perturbations in the non-CDM components as well. By $z_{init} = 200$, the linear perturbations in photons and neutrinos are completely subdominant on the scales of interest (see Appendix \ref{sec:ICsfromCAMB}). However, at late times massive neutrinos can cluster nonlinearly in the collapsing density perturbations so we allow for a contribution from the clustering of massive neutrinos given by
\be
\delta M_\nu (<R, t) =  \int_{V_R} d^3r \,\delta\rho_\nu(r,t)\,
\ee
where $V_R =  \frac{4}{3}\pi R^3$.

The final expression that we use to solve for for the subhorizon, non-linear evolution of $R(t)$ is then, 
\be
\label{eq:ddotR}
\ddot{R}=-\frac{GM}{R^2}-\frac{4\pi G}{3} \left(2\bar{\rho}_r(t)+\bar{\rho}_\nu(t) + 3\bar{P}_\nu-2\bar{\rho}_\Lambda(t)\right)R -\frac{ G \delta M_\nu(<R, t) }{R^2}\,,
\ee
with the initial conditions in Eq.~(\ref{eq:Rics}). 

Before proceeding it is helpful to rewrite Eq.~(\ref{eq:ddotR}) as a nonlinear equation for the density perturbation $\delta_{cb}$. Using $M = \frac{4}{3}\pi R^3\bar\rho_{cb} (1+\delta_{cb}) = const$ we have
\be
\label{eq:ddotdelta}
\ddot{\delta}_{cb} + 2H\dot{\delta}_{cb} -\frac{4}{3}\frac{\dot{\delta}_{cb}^2}{1+\delta_{cb}}-\frac{3}{2}H^2\left(\Omega_{cb}\delta_{cb}+\Omega_\nu\delta_\nu\right)(1+\delta_{cb}) = 0
\ee
where $\Omega_{cb} = \Omega_c(a) + \Omega_b(a)$ and $\delta_\nu = \delta M_\nu/\bar{M}_\nu$. Collapse of a halo occurs when $R\rightarrow 0$ or equivalently $\delta_{cb}\rightarrow \infty$. In the absence of neutrino perturbations $\delta_\nu$ (which depend on halo mass $M$) the equation of motion for $\delta_{cb}$ Eq.~(\ref{eq:ddotdelta}) is completely independent of halo mass. Therefore, in the absence of neutrino clustering a given set of initial conditions $\delta_{cb,init}$, $\dot\delta_{cb,init}$ will collapse at the same time, regardless of the halo mass. The value of $\dot\delta_{cb}(z_{init})/\delta_{cb}(z_{init})$, however, is cosmology dependent and there are percent-level changes for the different choices of $m_{\nu}$ and $\Omega_m$ (see Appendix \ref{sec:ICsfromCAMB}). In this paper we are particularly interested in the importance of the final $\delta M_\nu$ term. Calculations that ignore the $\delta M_\nu$ term are referred to as neglecting neutrino clustering while those that incorporate a non-zero $\delta M_\nu$ are said to include neutrino clustering. In the next section we outline our methods for calculating $\delta M_\nu$. 

\subsection{Expression for $\delta M_\nu(<R, t)$}
\label{ssec:deltarhonu}

\begin{figure}[t]
\begin{center}
$\begin{array}{cc}
 \includegraphics[width=0.5\textwidth]{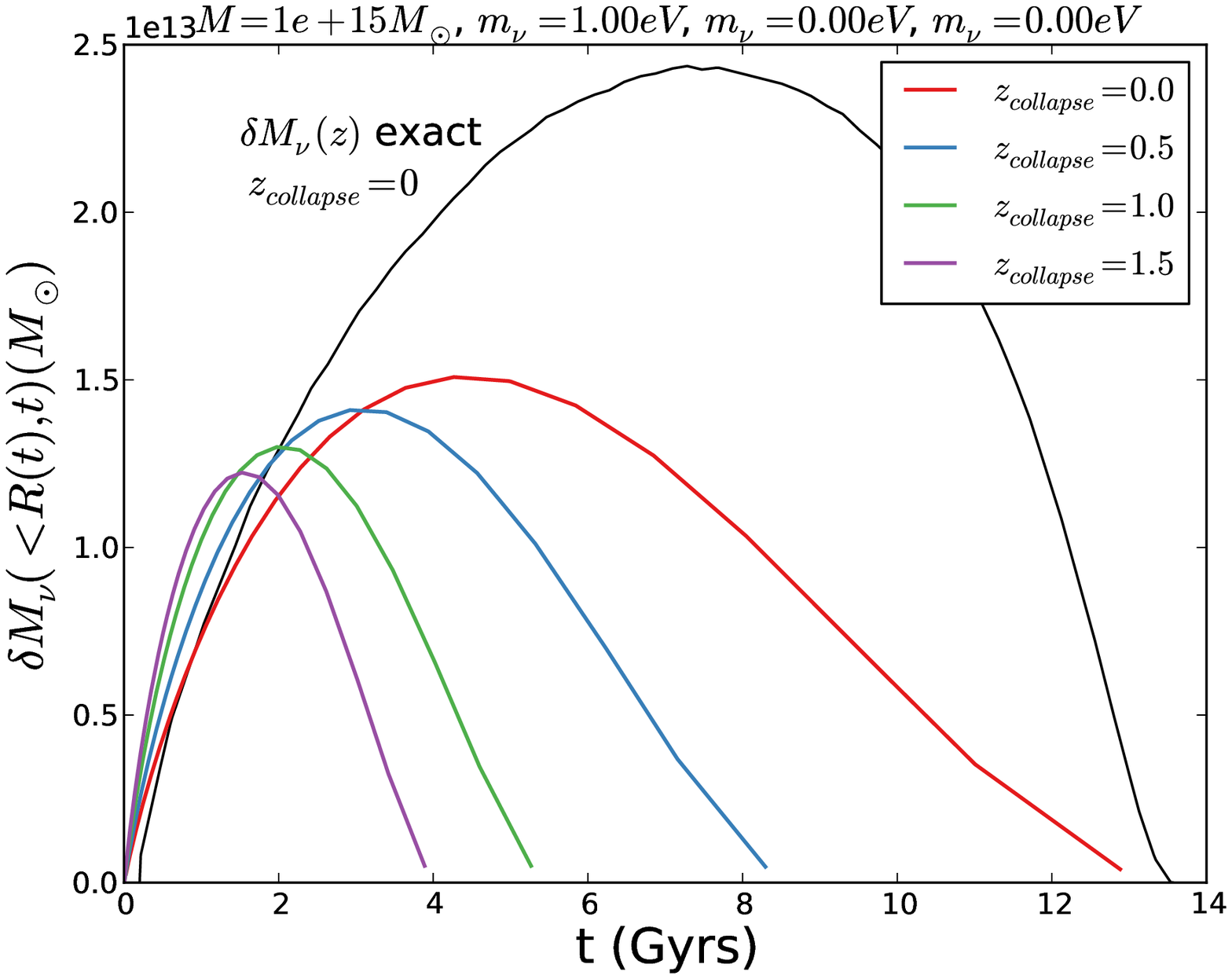} & \includegraphics[width=0.5\textwidth]{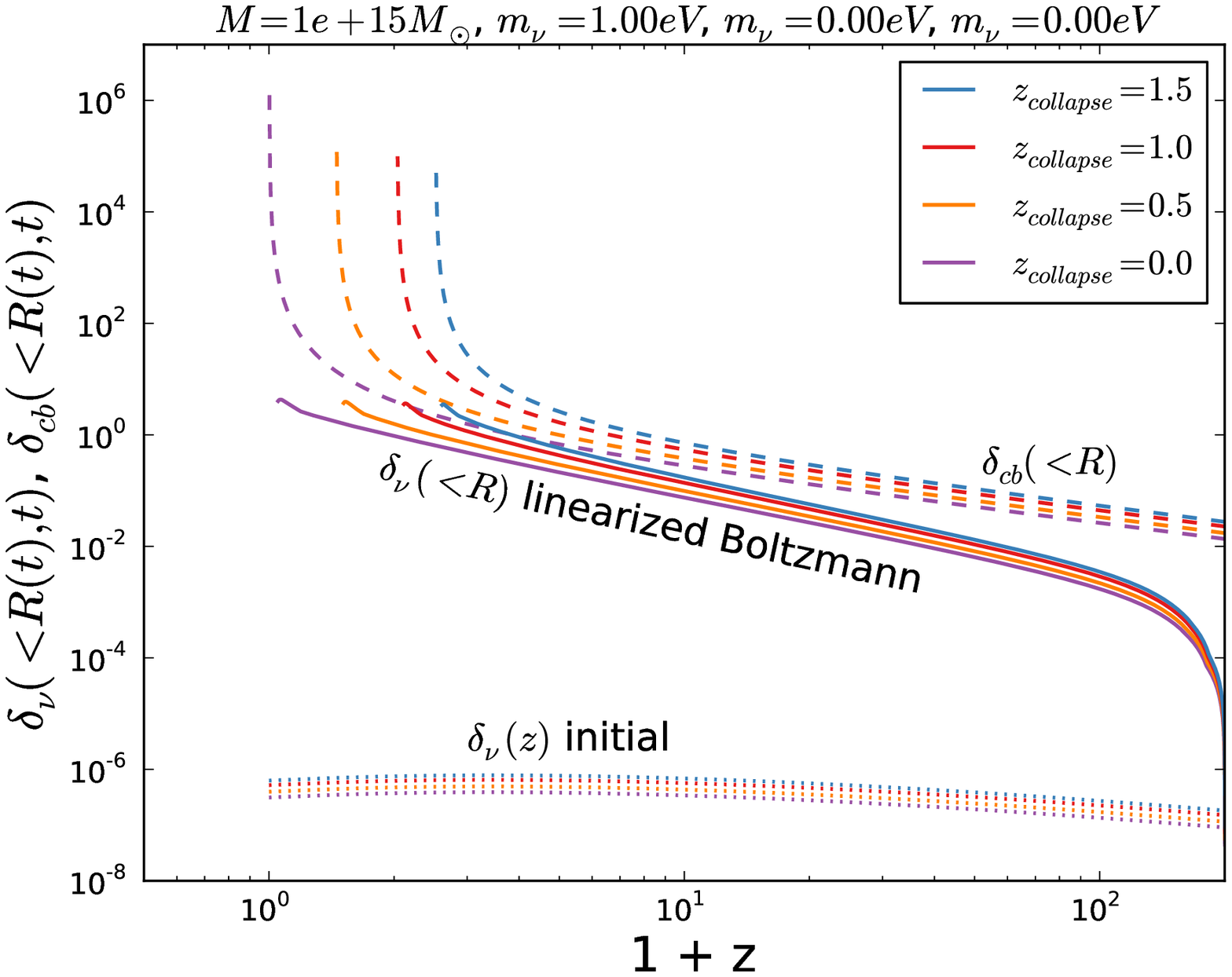}\\
 \mbox{(a)} &  \mbox{(b)} 
\end{array}$
\caption{\label{fig:deltaMnu} Left panel: The total neutrino mass fluctuation within radius $R$ for neutrinos with $m_\nu = 1eV$, calculated using the linearized Boltzmann equation (Eqs.~(\ref{eq:deltaMnu}) and ~(\ref{eq:f1})) for halos of $M = 10^{15} M_\odot$ that collapse at several different times. For the halo that collapses latest, we have also plotted $\delta M_{\nu}$ calculated using the exact calculation (see Appendix \S \ref{sec:deltaMnutests} for details).  Right panel: The neutrino density fluctuation interior to $R$, $\delta M_\nu/\bar{M}_\nu$ (solid lines), compared with the CDM + baryon density fluctuation (dashed lines), and the initial linear neutrino fluctuation from CAMB (dotted lines). The divergences in the density fluctuations correspond to the redshift at which $R\rightarrow 0$.}
\end{center}
\end{figure}

The neutrino mass perturbation interior to $R$, $\delta M_\nu$, is calculated by treating the halo as an external potential for the neutrinos. We use the no-neutrino-clustering solution (that is, the solution to Eq.~(\ref{eq:ddotR}) with $\delta M_\nu = 0$) for $R(t)$ to determine the potential due to the CDM + baryon density perturbation. We then use this potential as input into the linearized Boltzmann equation for the evolution of the first-order perturbation to the neutrino distribution function (see  \cite{1966ApJ...144..233G, Brandenberger:1987kf,Singh:2002de}). 

The neutrino mass fluctuation interior to $R$ at time $t$ from a single species of mass $m_\nu$ is given by
\be
\label{eq:deltaMnu}
\delta M_\nu (< R, t) =  m_\nu \int_{V_c} d^3 {\bf r} \int \frac{d^3{\bf q}}{(2\pi)^3}f_1({\bf q}, {\bf r}, t)
\ee
where ${\bf q} = a{\bf p}$ and ${\bf p}$ is the particle momentum, $f_1$ is a linear perturbation to the neutrino distribution function ($f= f_0 + f_1$ where $f_0$ is a Fermi-Dirac distribution with temperature $T_\nu$) and $V_c \equiv \frac{4}{3}\pi R^3(t)/a^3(t)$. 

For $f_1$, we use an approximate solution to the Boltzmann equation for non-relativistic particles in an external potential (for details see \cite{LoVerde:2013lta}). For neutrinos around a spherical top-hat halo, this reads
\ba
\label{eq:f1}
f_1({\bf r}_{comoving},{\bf q}, \eta) &=& 2\frac{m_\nu}{T_\nu} \int_{t_0}^t \frac{dt'}{ a(t')} \frac{ e^{q/T_\nu}}{(e^{q/T_\nu}+1)^2}\frac{G\delta M(t')}{r^2}\left(\alpha \frac{q}{T_\nu}- \hat{q}\cdot\hat{r}\right)\\
&&\left\{\frac{a^3(t')r^3}{R^3}\Theta\left(r^2(1+q^2/T_\nu^2\alpha^2 -2 q/T_\nu\alpha\hat{r}\cdot\hat{q}) < R(t')^2/a^2(t')\right) \right.\nn\\
&&\left.+ \frac{\Theta\left(r^2(1+q^2/T_\nu^2\alpha^2 -2 q/T_\nu\alpha\hat{r}\cdot\hat{q}) \ge R(t')^2/a^2(t')\right) }{\left(1+q^2/T_\nu^2\alpha^2 -2 q/T_\nu\alpha\hat{r}\cdot\hat{q} \right)^{3/2}}\right\}\nn
\ea
where $\delta M(t) = M - \frac{4}{3}\pi \bar{\rho}_{cb}R^3$, $\Theta$ is the Heaviside step function, $\alpha \equiv  {T_\nu(\eta-\eta')}/{m_\nu r}$, and $\eta$ is a time variable defined by $a^2 d\eta = dt$. Figure \ref{fig:deltaMnu} shows $\delta M_\nu$ along with the fractional overdensity $\delta M_\nu/\bar{M}_\nu$ for a single neutrino species with $m_{\nu} = 1eV$ in halos of $M=10^{14} M_\odot$ collapsing at several different times.  

The linearized solution in Eq.~(\ref{eq:f1}) underestimates the neutrino clustering on scales interior to $R$ \cite{Ringwald:2004np, LoVerde:2013lta}; this can be seen explicitly in Fig. \ref{fig:deltaMnu} where we have also plotted an example calculation of $\delta M_\nu$ from a full non-linear solution to the Boltzmann Equation for $m_\nu = 1eV$.  The difference between the linear approximation for $\delta M_\nu$ and the exact solution is large, particularly at late times. However, we found in \cite{LoVerde:2013lta} that for $m_\nu \lsim 0.2eV$, significant differences between the linearized and exact calculations of the neutrino mass in an external potential do not become important until after the halo has begun to collapse; at this point the cold dark matter overdensity is large and the dynamics of $R$ are dominated by the CDM. As shown in Appendix \S \ref{sec:deltaMnutests}, even for more massive neutrinos ($m_{\nu} = 1eV$) the linearized solution is sufficient to determine the evolution of $R$ and the collapse time to about a percent. The fractional differences between the values of $t_{collapse}$ for a given $\delta_{cb,i}$  using the linearized Boltzmann solution and using the exact solution for $\delta M_\nu$ remain about an order of magnitude smaller than the fractional differences between $t_{collapse}(\delta M_{\nu})$ and $t_{collapse}(\delta M_{\nu} = 0)$. 

In panel $(b)$ of Fig. \ref{fig:deltaMnu} we compare the the fractional overdensity in neutrinos accumulating in the halo (calculated using the linearized Boltzmann solution), the fractional overdensity in CDM and baryons, and the initial linear density fluctuation in neutrinos from the CAMB transfer functions. On halo scales, the initial linear fluctuation in the neutrino density has damped away and is irrelevant in comparison to the neutrino mass that accumulates in the halo at later times. 

\begin{figure}[t]
\begin{center}
$\begin{array}{cc}
 \includegraphics[width=0.5\textwidth]{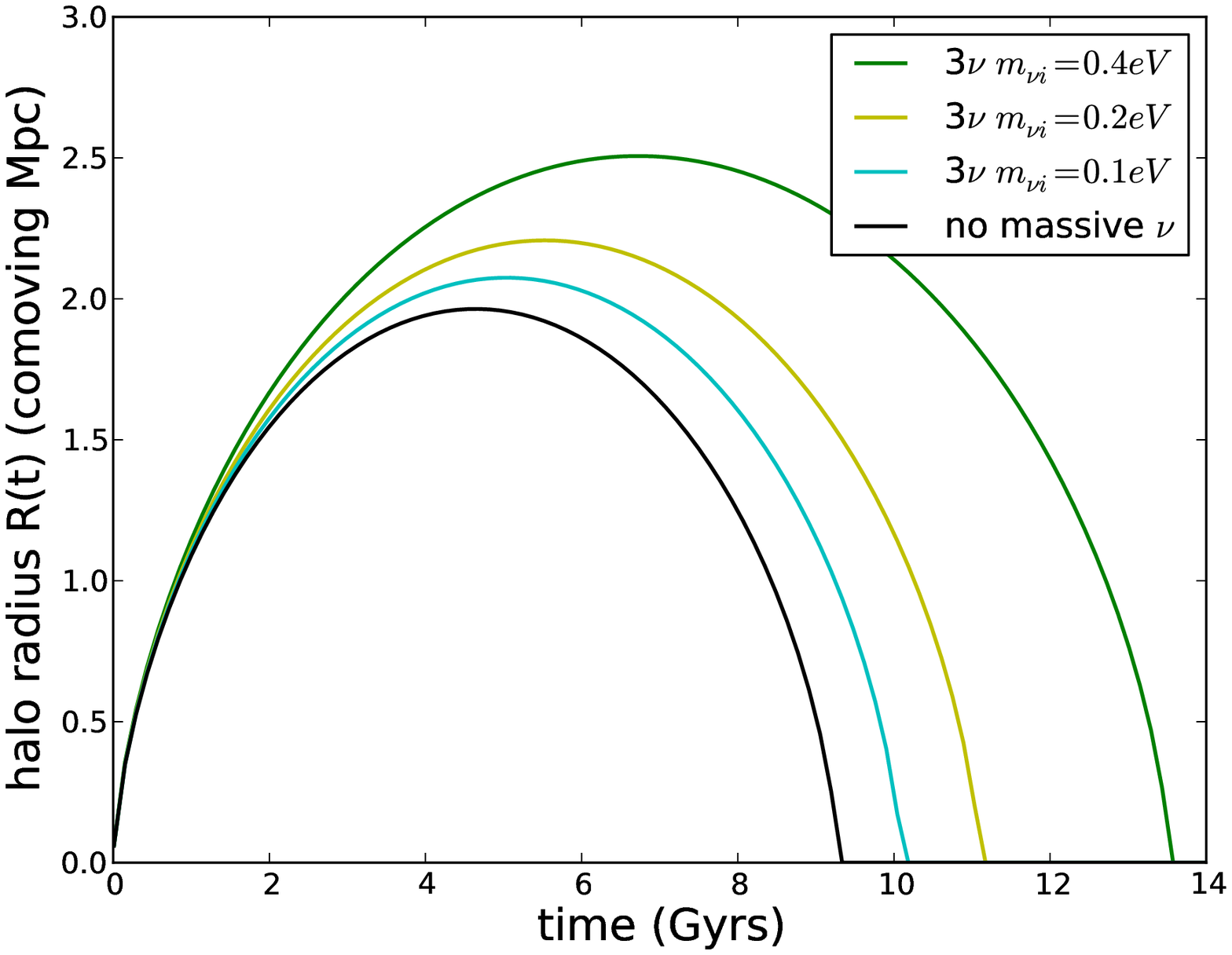} & \includegraphics[width=0.5\textwidth]{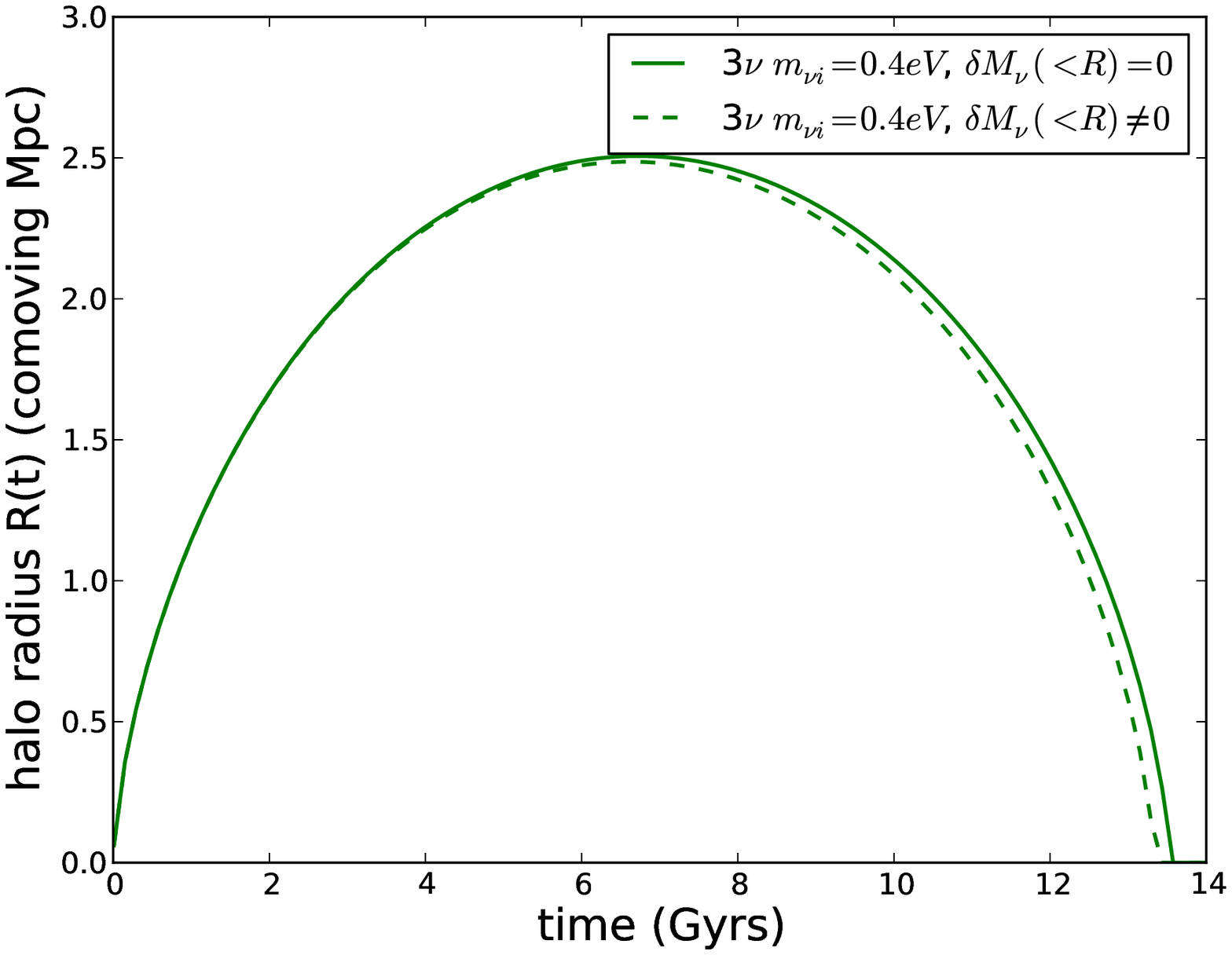}\\
  \mbox{(a)} &  \mbox{(b)} 
\end{array}$
\caption{\label{fig:Roft} Left panel: Spherical collapse solutions for halos with common values of $\delta_{init}$ but in cosmologies with different $m_{\nu}$; each curve above has $m_{\nu 1} = m_{\nu 2} = m_{\nu 3}$ and from bottom to top they are $m_{\nu i} =0\,eV$, $0.1\,eV$, $0.2\,eV$, $0.4\,eV$. These curves neglect neutrino clustering ($\delta M_\nu(<R) =0$). In cosmologies with massive neutrinos the linear growth is slower and collapse occurs later. Right: A comparison of the spherical collapse solutions including and neglecting neutrino clustering interior to $R$ for a cosmology with $m_{\nu i} = 0.4\, eV$.}
\end{center}
\end{figure}

In panel $(a)$ of Fig.~\ref{fig:Roft} we plot solutions to Eq.~(\ref{eq:ddotR}) with $\delta M_{\nu}(<R) =0$ for halos with the same initial $\delta_{cb}(z_{init})$ (and therefore the same $R_{init}$) but in cosmologies with different neutrino masses so the $\dot{\delta}_{cb}(z_{init})$ and the background evolution differ. Increasing the neutrino mass delays the evolution and subsequent collapse of the halos. In panel $(b)$ we show the effect of neutrino clustering on the evolution of $R(t)$. As expected, neutrino clustering interior to $R$ causes a slight decrease in the collapse time. 

\label{sec:sphercollapseresults}

\section{From Spherical Collapse to Halo Abundance}
\label{sec:massfcn}
In this section we relate the spherical collapse calculations from the previous section to the abundance of halos at late times. As we will discuss below, the scale-dependent evolution of density perturbations in $\nu \Lambda CDM$ introduces some subtleties that are not present in the $CDM$-only case so we review the spherical collapse framework for halo abundance and our assumptions in detail. 

\subsection{The critical overdensity}
\label{ssec:deltacrit}
In the spherical collapse model, one approximates the nonlinear evolution of the initial density field smoothed on scale $R$ with the evolution of a spherical top-hat density perturbation with the same initial amplitude $\delta_{cb}(z_{init})$ and velocity $\dot\delta_{cb}(z_{init})$. From the solution to the equation of motion for $R(t)$ one can determine how large $\delta_{cb}(z_{init})$ and $\dot\delta_{cb}(z_{init})$ need to be for a halo to have collapsed ($R\rightarrow 0$) by a given redshift. The critical value of $\delta_{cb}(z_{init})$ required for a spherical halo to collapse by redshift $z$ is called the collapse threshold or critical overdensity, $\delta_{crit}(z_{init})$, and is a key ingredient in analytic models of the halo abundance (see for instance, \cite{Press:1973iz, Bardeen:1985tr}). In the next few paragraphs we discuss our approach for determining $\delta_{crit}(z_{init})$.

The initial amplitude of the smoothed density field around a point $\x$ is given by 
\be
\label{eq:deltaR}
\delta_{cb,R}(\x,z_{init}) = \int \frac{d^3\k}{(2\pi)^3} e^{i\k\cdot\x} W(kR)\delta_{cb}(\k,z_{init})
\ee
and the initial velocity of the pertrubation is
\be
\label{eq:dotdeltaR}
\dot\delta_{cb,R}(\x,z_{init}) = \int \frac{d^3\k}{(2\pi)^3} e^{i\k\cdot\x} W(kR)\frac{d\ln \delta_{cb}}{dt}(\k,z_{init})\delta_{cb}(\k,z_{init})\,.
\ee
The initial perturbation amplitude $\delta_{cb,R}(\x,z_{init})$ is a Gaussian random variable with zero mean and variance  
\be
\label{eq:sigma2M}
\sigma^2(M,z) \equiv \langle \delta_R^2\rangle = \int \frac{dk}{k} \left|W(kR)\right|^2 \frac{k^3 P_{cb}(k,z)}{2\pi^2} 
\ee
evaluated at $z = z_{init}$, where $P_{cb}(k,z)$ is the power spectrum of cold dark matter plus baryon perturbations and 
\be
\label{eq:WkR}
W(k,R) = \frac{3 j_1(kR(M)}{kR(M)},
\ee
where $j_1(x)$ is the  the spherical Bessel function and $R(M) = (3M/(4\pi\bar\rho_{cb}))^{1/3}$. The variance of the velocity of the initial perturbations is
\be
\label{eq:dotsigma2M}
\dot\sigma^2(M,z_{init}) \equiv \langle \dot\delta_R^2\rangle = \int \frac{dk}{k} \left|W(kR)\right|^2\left(\frac{d\ln \delta_{cb}}{dt}(k,z_{init})\right)^2 \frac{k^3 P_{cb}(k,z_{init})}{2\pi^2}\,. 
\ee
where $d\ln \delta/dt(k,z_{init})$ is the linear evolution of the growing mode (as determined by CAMB, for instance). 

If the linear evolution is scale independent $d\ln \delta_{cb}/dt$ can be pulled out of the integrals in Eq.~(\ref{eq:dotdeltaR}) so that $\dot{\delta}_{cb,R}(\x,z_{init})$ is proportional to $\delta_{cb,R}(\x,z_{init})$ regardless of the density profile around $\x$. That is, for scale-independent evolution the value of $\delta_{cb,R}(\x, z_{init})$ alone (rather than the full profile $\delta_{cb}(\k,z_{init})$) completely specifies $\dot{\delta}_{cb,R}(\x, z_{init})$. On the other hand, in a cosmology with scale-dependent growth (such as the case with massive neutrinos) $d\ln \delta_{cb}/dt(k)$ can not be factored out of Eq.~(\ref{eq:dotdeltaR}) and the value of $\delta_{cb,R}(\x,z_{init})$ alone does not determine $\dot{\delta}_{cb,R}(\x,z_{init})$. 

To determine the collapse threshold for a halo of mass $M$ we solve Eq.~(\ref{eq:ddotR}) for the evolution of $R$ for a top-hat perturbation with some initial amplitude $\delta_{cb,init}$ and initial velocity 
\be
\label{eq:sigmadeltadot}
\dot\delta_{cb,init} \equiv \frac{\dot{\sigma}(M,z_{init})}{\sigma(M,z_{init})}\delta_{cb,init}\,.
\ee
We then determine how large $\delta_{cb}(z_{init})$ needs to be for the perturbation to collapse by a given redshift $z$. This defines the critical overdensity at the initial time $\delta_{crit,init}$. With this definition the critical overdensity linearly extrapolated to a different redshift is
\be
\label{eq:deltacrit}
\delta_{crit}(z) \equiv \frac{\sigma(M,z)}{\sigma(M,z_{init})}\delta_{crit, init}\,.
\ee
We have chosen Eq.~(\ref{eq:sigmadeltadot}) and Eq.~(\ref{eq:deltacrit}) to define the collapse threshold because Eq.~(\ref{eq:sigmadeltadot}) is representative of {\em typical} initial conditions for the quantity $\delta_{cb,R}(\x,z_{init})$. Moreover, from Eq.~(\ref{eq:deltacrit}) we can see that the rarity of perturbations large enough to collapse as characterized by the ratio $\delta_{crit}(z)/\sigma(M,z)$ is independent of redshift. In contrast, the linear evolution of an initial  density perturbation with an exact top-hat density profile, $\delta^{top-hat}(k, z_{init}) = 4\pi/(2\pi)^3 k^2 W(kR)$ differs from the linear evolution of $\sigma(M,z)$ if $d\ln\delta/dt(k)$ is $k$-dependent because $\delta^{top-hat}$ and $\sigma(M,z)$ depend differently on $k$.  In Appendix \ref{sec:comparedeltacrit} we discuss this issue in detail and make comparisons between the definition used here and some alternative approaches. 

\begin{figure}[t]
\begin{center}
$\begin{array}{cc}
 \includegraphics[width=0.5\textwidth]{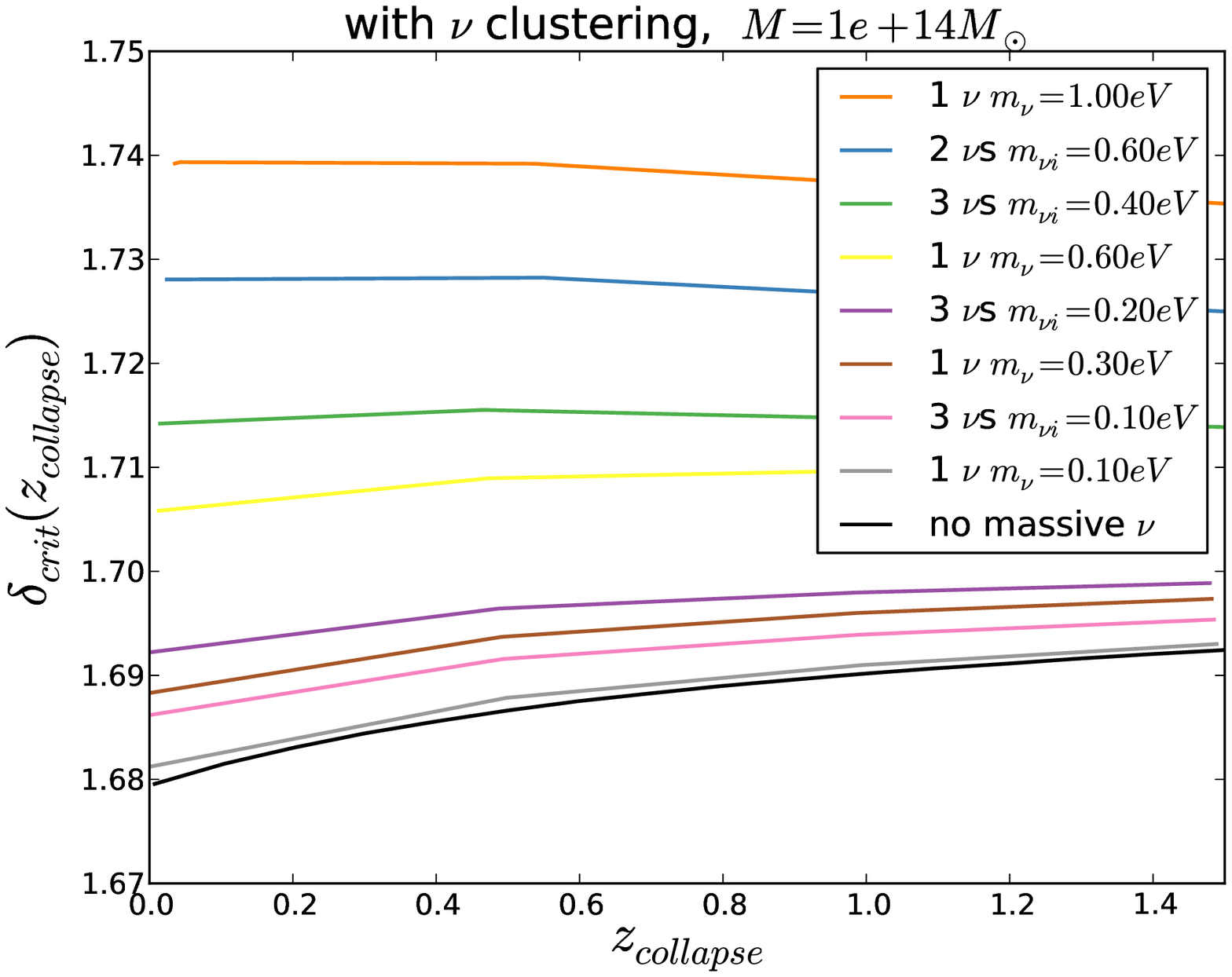}&   \includegraphics[width=0.5\textwidth]{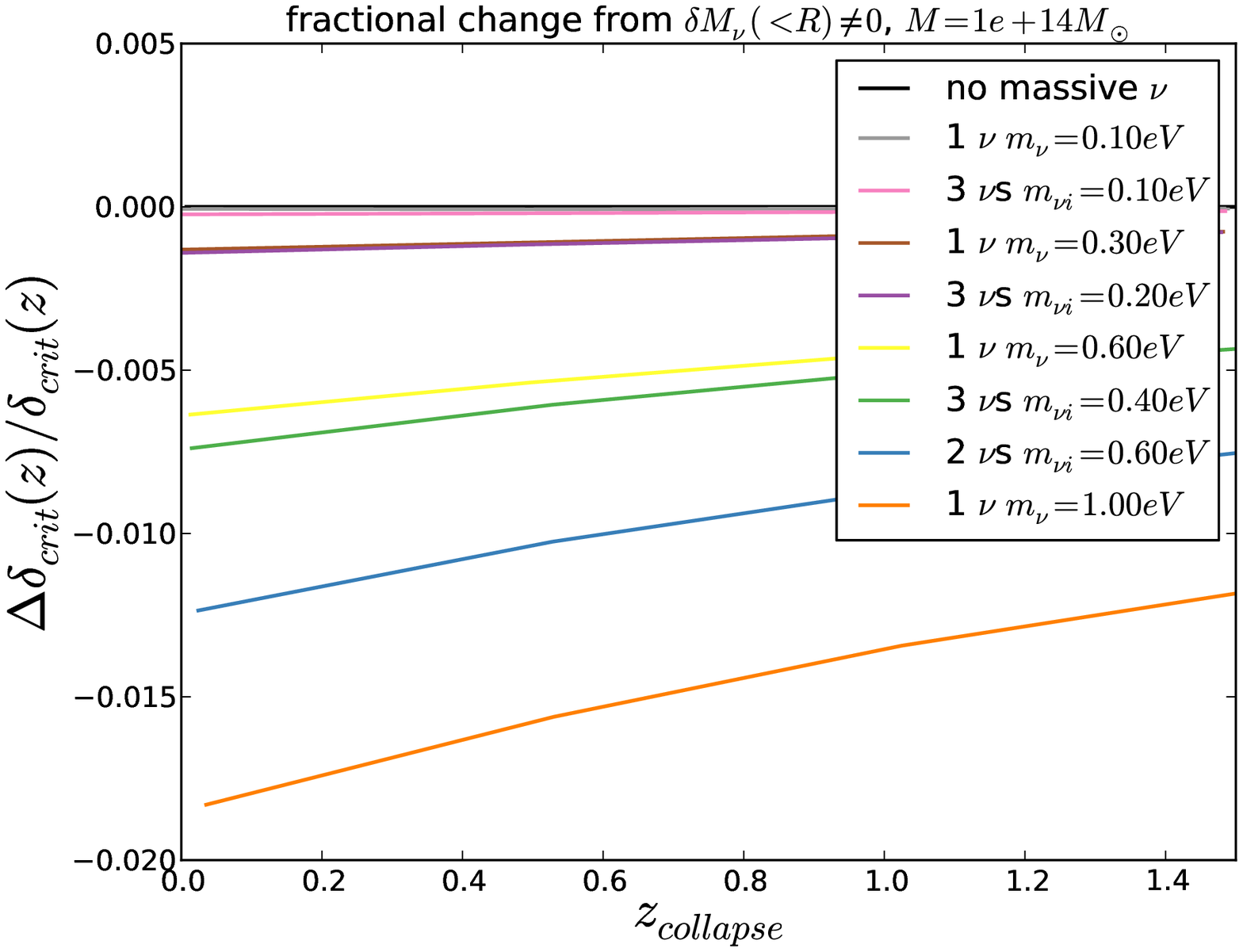} \\
 \mbox{(a)} & \mbox{(b)}
\end{array}$
\caption{\label{fig:deltac} Left: The linearly extrapolated value of the initial density perturbation $\delta_{cb}(z_{init})$, required to collapse at $z_{collapse}$ (see Eq.~(\ref{eq:deltacrit})). Here we have fixed $\Omega_c$ and $\Omega_b$ so curves with different $\Omega_\nu$ have different total matter density $\Omega_m$. Right: The fractional change to the collapse threshold when neutrino clustering interior to $R$ is included. In both panels $M = 10^{14} M_\odot$ and the order of the curves matches the order of the legends. }
\end{center}
\end{figure}

In Fig. \ref{fig:deltac} we plot the collapse threshold linearly extrapolated to the collapse redshift, $\delta_{crit}(z_{collapse})$, for a range of neutrino mass hierarchies. Increasing the total amount of neutrino mass increases the barrier for collapse, but in a way that depends on the individual neutrino masses (as opposed to just $\Omega_\nu \propto \sum_i m_{\nu i}$). For instance, the scenario $m_{\nu 1} = 0.6eV $, $m_{\nu 2} = m_{\nu 3} = 0eV$ and the degenerate hierarchy $m_{\nu i} = 0.2eV$ have common values of $\sum_im_{\nu i}$ (and therefore $\Omega_{\nu}$ once the neutrinos are nonrelativistic) but the $\delta_{crit}(z)$'s clearly differ. The non-linear dependence on $m_{\nu}$ is partially due to the nonlinear clustering of massive neutrinos during collapse (roughly $\delta M_{\nu} \sim \sum_i m_{\nu i}^{5/2}$ \cite{LoVerde:2013lta}), which is shown in panel (b) of Fig. \ref{fig:deltac}. However, even in the absence of nonlinear neutrino clustering $\delta_{crit}$ has some sensitivity to the individual neutrino masses through the suppression in linear growth:  the net suppression in small-scale density perturbations depends on both the fraction of mass in neutrinos and the redshift at which the neutrinos become nonrelativistic ($a_{NR} \approx 3T_\nu /m_\nu$) which gives an additional sensitivity to the masses \cite{Hu:1997mj}.

\begin{figure}
\begin{center}
$\begin{array}{cc}
%\includegraphics[width=0.5\textwidth]{plots_camb/Deltadeltac_vs_nbar_mnu00_M1e+14.eps} & \includegraphics[width=0.5\textwidth]{plots_camb/Deltadeltac_vs_mnu_mnu00_M1e+14.eps}\\
%\mbox{(a)} & \mbox{(b)}\\
\includegraphics[width=0.5\textwidth]{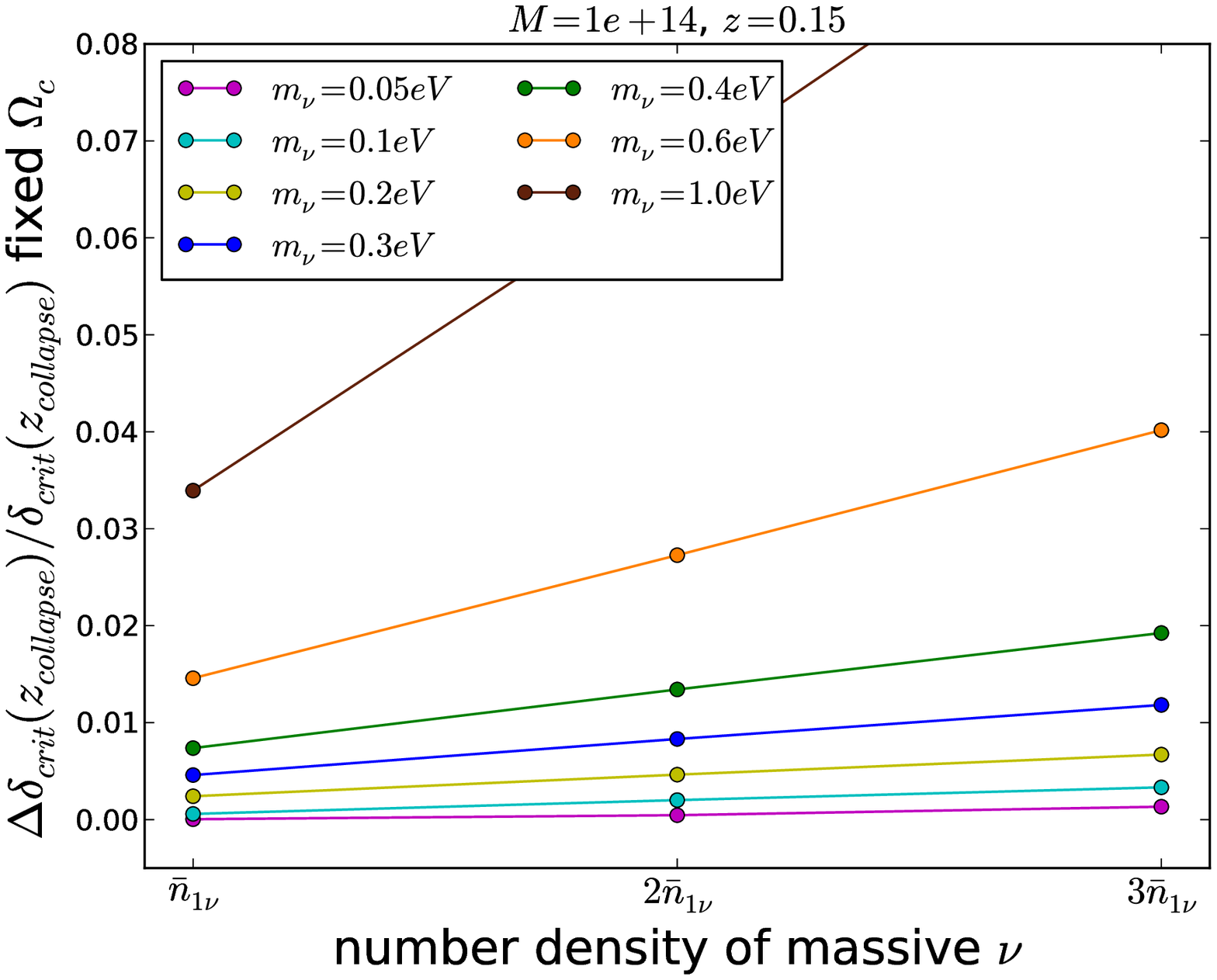} & \includegraphics[width=0.5\textwidth]{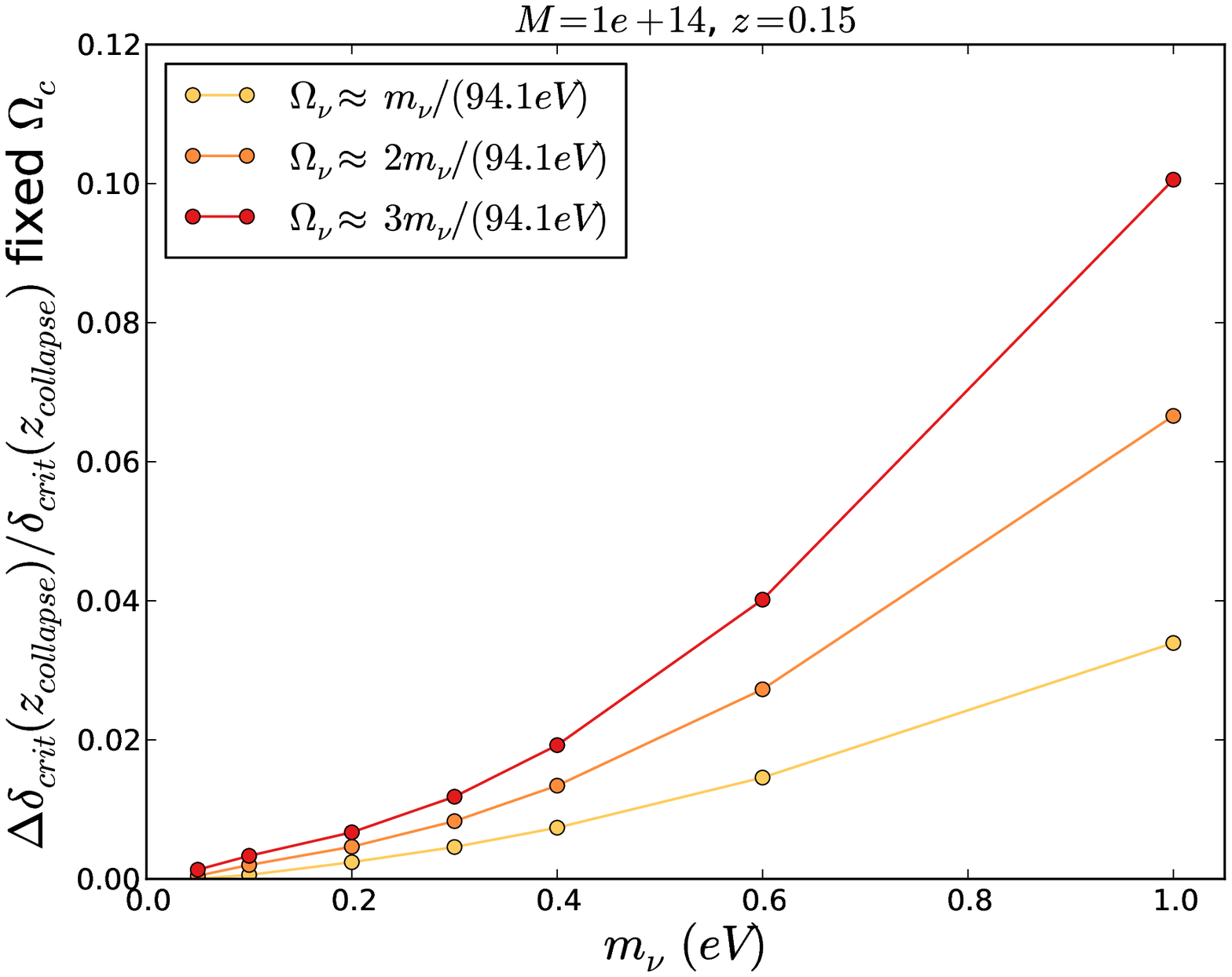}\\
\mbox{(a)} & \mbox{(b)}
\end{array}$
\caption{\label{fig:delta_clustering_mndep} The change to $\delta_{crit}(z_{collapse})$ due to massive neutrinos (including neutrino clustering and compared at fixed $\Omega_c$). Left: The dependence on the number density of massive neutrinos, plotted for several different neutrino masses in units of $\bar{n}_{1\nu}$, i.e., the relic abundance of a single neutrino and antineutrino species given in Eq.~(\ref{eq:nbarnu}). The change to the collapse threshold is roughly linear in $\bar{n}_{1\nu}$.  Right: The change to the linearly extrapolated collapse threshold plotted as a function of neutrino mass $m_{\nu}$. The change to $\delta_{crit}(z_{collapse})$ depends non-linearly on the masses of the individual neutrinos. }
\end{center}
\end{figure}

In Fig.~\ref{fig:delta_clustering_mndep} we show the dependence of the change to the collapse threshold $\delta_{crit}(z_{collapse})$ on the abundance of relic neutrinos and their individual masses. For the range of neutrino masses and halo masses we have considered ($10^{14} M_\odot < M < 10^{16}M_\odot$ and $0.05eV < m_{\nu i } < 1eV$) increasing $\bar{n}_{\nu}$ causes a linear increase in $\delta_{crit}$.  On the other hand,  $\delta_{crit}(z_{collapse})$ depends non-linearly on the masses of the individual neutrinos. This means that observables dependent on $\delta_{crit}(z_{collapse})$ such as the halo mass function are in principle able to distinguish between different scenarios for the neutrino mass hierarchy.  In practice, the non-linear dependence of $\delta_{crit}(z_{collapse})$ on the neutrino masses is probably only important for scenarios with an additional sterile species but it is nevertheless important to keep in mind. 


\subsection{The halo abundance}
The calculations of the collapse threshold from \S \ref{ssec:deltacrit} can be used to estimate the effects of massive neutrinos on the halo mass function. To convert the changes to $\delta_{crit}(z_{collapse})$ and $\sigma(M,z)$ into predictions for changes to the halo abundance we will need to make the assumption that the changes to the halo mass function are characterized entirely by these two parameters, an assumption that will need to be tested against N-body simulations with neutrino masses across the ranges that we consider.

To study neutrino mass effects on the halo mass function, we use the fitting formula of Bhattacharya et al \cite{Bhattacharya:2010wy} which is calibrated off of high-resolution CDM-only N-body simulations. Their expression for the number density of halos with masses between $M$ and $M+dM$ is
\be
\label{eq:nB}
n_B(M,z) = -\frac{\bar{\rho}_{cb}}{M} \frac{d\ln \sigma(M,z)}{dM} f_B\left(\nu \equiv \frac{\delta_{crit}(z)}{\sigma(M,z)},z\right)\,,
\ee
where 
\be
\label{eq:fB}
f_B(\frac{\delta_{crit}}{\sigma}, z) =  A\sqrt{\frac{2}{\pi}} \exp\left\{-\frac{a}{2}\frac{\delta_{crit}^2}{\sigma^2}\right\}\left(1 + \left(\frac{\sigma^2}{a\delta_{crit}^2}\right)^{p}\right)\left(\frac{\delta_{crit}\sqrt{a}}{\sigma}\right)^q\,,
\ee
with $A(z) = 0.333/(1+z)^{0.11}$, $a = 0.788/(1+z)^{0.01}$, $p = 0.807$, $q = 1.795$, and $\sigma = \sigma(M,z)$. This fitting function was calibrated off of N-body simulations of $\Lambda CDM$ using a constant linearly extrapolated spherical collapse threshold $\delta_{crit}(z) = 1.686$. To study the effects of massive neutrinos on halo abundance we will therefore use 
\be
\label{eq:deltacritsub}
\delta_{crit}\rightarrow 1.686\frac{\delta_{crit}(z|m_{\nu})}{\delta_{crit}(z|m_{\nu} =0)}
\ee
in Eq.~(\ref{eq:nB}) where $\delta_{crit}(z |m_{\nu})$ and $\delta_{crit}(z|m_{\nu}=0)$ are our calculated collapse thresholds from Eq.~(\ref{eq:deltacrit}) \footnote{We have checked that using $\delta_{crit}(z |m_{\nu})$ directly, rather than the replacement in Eq.~(\ref{eq:deltacritsub}), makes only a small difference in the predicted change to the mass function. Precisely, the difference is $\lsim 10\%$ for $M \lsim 10^{15}h^{-1}M_\odot$ and it is small compared to the total change in $n(M)$ due to $m_{\nu} \neq 0$.} . We use $\sigma(M,z)$ from Eq.~(\ref{eq:sigma2M}) including the effects of massive neutrinos on the linear CDM and baryon power spectrum. 

For comparison, we will also consider the predicted changes to the Sheth-Tormen (ST) mass function
\be
n_{ST}(M,z) = -\sqrt{\frac{2q_{ST}}{\pi}}A_{ST}\left(1+\left(\frac{q_{ST}\delta_{crit}^2}{\sigma^2}\right)^{-p_{ST}}\right)\frac{\bar{\rho}_{cb}}{M^2}\frac{\delta_{crit}}{\sigma}\frac{d\ln \sigma}{d\ln M}\exp\left(-\frac{q_{ST}\delta^2_{crit}}{\sigma^2}\right)
\ee
where $q_{ST} = \sqrt{2}/2$, $A_{ST} = 0.322184$, and $p_{ST} =0.3$. 

\begin{figure}[t]
\begin{center}
$\begin{array}{cc}
 \includegraphics[width=0.5\textwidth]{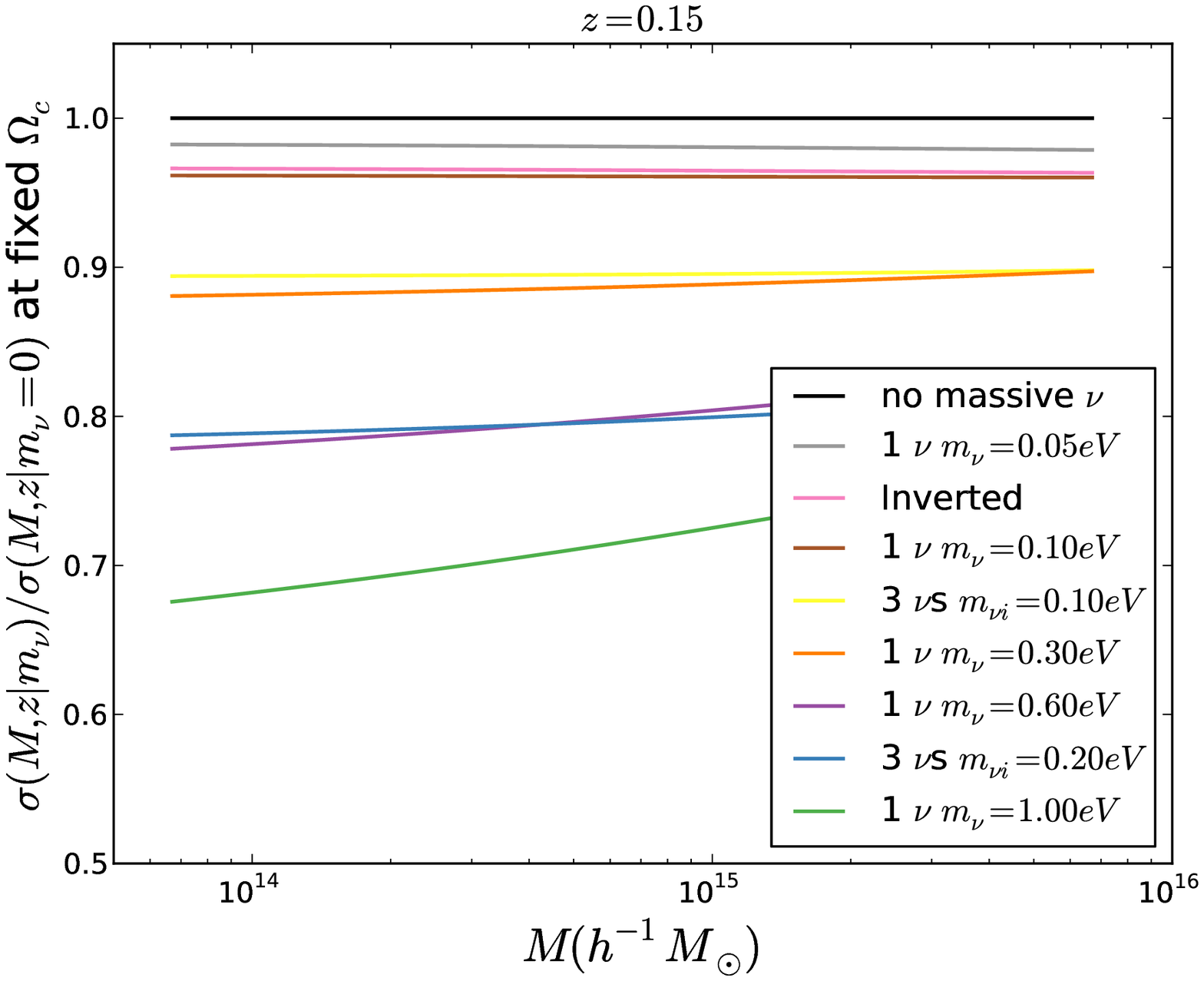} & \includegraphics[width=0.5\textwidth]{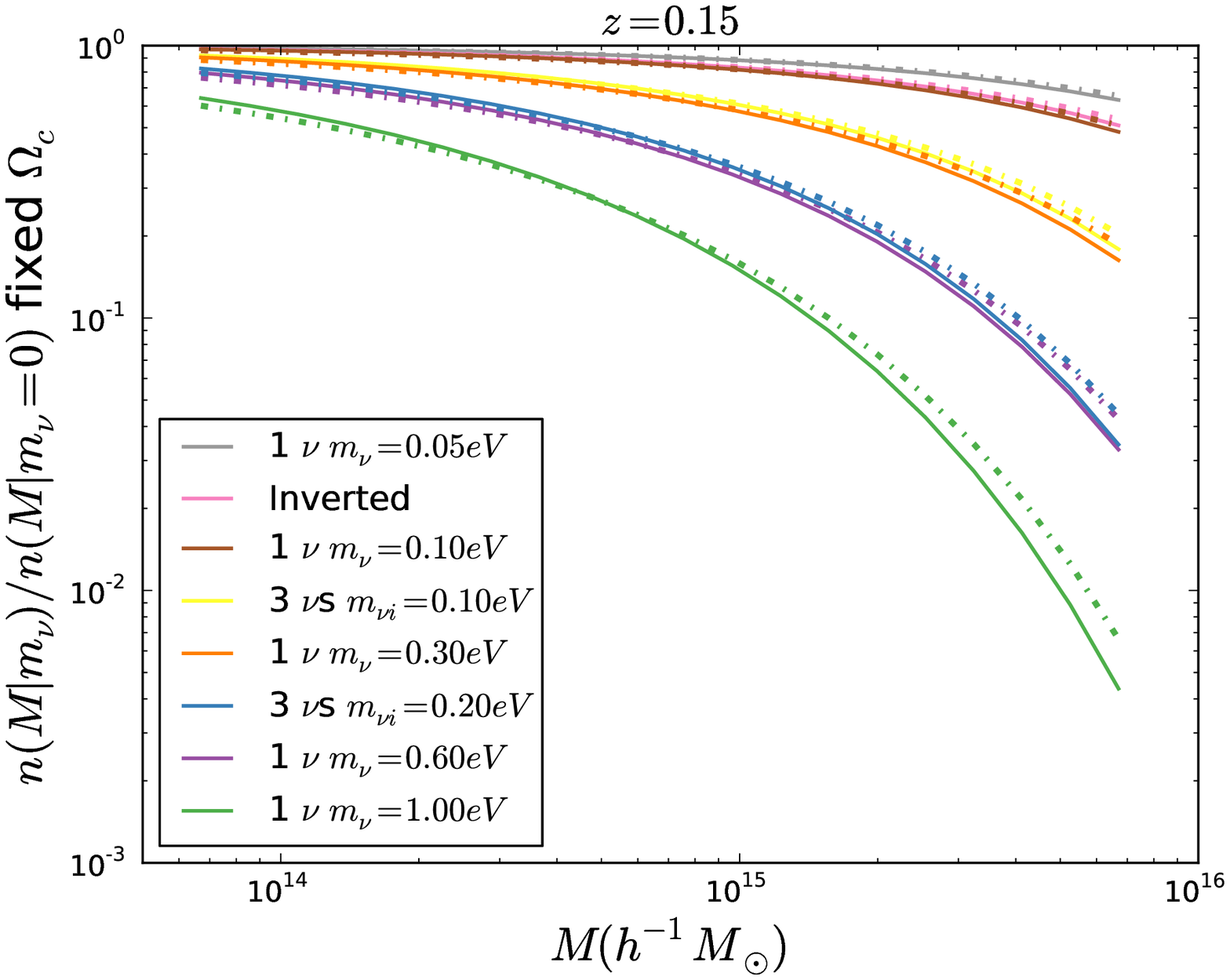}\\
  \mbox{(a)} &  \mbox{(b)} 
\end{array}$
\caption{\label{fig:deltanMs}  Left: The  neutrino-induced suppression in the amplitude of $\sigma(M)$ -- the CDM and baryon density fluctuations smoothed on scale $M$ (Eq.~(\ref{eq:sigma2M})). Right: The change to the halo abundance in a cosmology with massive neutrinos at $z=0.15$ using the corrections to $\delta_{crit}$ for $m_{\nu}\neq 0$ calculated in this paper and the mass functions of Bhattacharya (solid lines) and Sheth-Tormen (dot-dashed lines). In both panels the order of the legend matches the order of the curves.}
\end{center}
\end{figure}

In Fig.~\ref{fig:deltanMs}, we show the suppression in variance of CDM and baryon density fluctuations $\sigma(M)$ due to massive neutrinos (Eq.~(\ref{eq:sigma2M})).  The fractional change in $\sigma(M)$ due to massive neutrinos is large compared to the shifts in $\delta_{crit}$ shown in Figs.~\ref{fig:deltac} and \ref{fig:delta_clustering_mndep}. One can also see in panel (a) of Fig.~\ref{fig:deltanMs} that the suppression in $\sigma(M)$ is not entirely characterized by $\Omega_\nu$: there are small differences between curves with common $\Omega_\nu$ but different $m_{\nu i}$ and this causes $n(M)$ to retain some sensitivity to individual neutrino masses even if $\delta_{crit}$ is independent of the neutrino mass hierarchy. 

In panel (b) of Fig.~(\ref{fig:deltanMs}) we show the changes to the halo mass function for cosmologies with massive neutrinos. We show the fractional correction to the both the Bhattacharya mass function and the Sheth-Tormen mass function. The dominant changes to the mass function are from the shift in $\sigma(M)$ rather than $\delta_{crit}$ shown in Fig.~\ref{fig:deltacTH} and Fig.~\ref{fig:delta_clustering_mndep}. While scenarios with common $\Omega_\nu$ but different $m_{\nu i}$ are not completely degenerate,  the differences are small and it would be very challenging to distinguish between them.  

\begin{figure}[t]
\begin{center}
 \includegraphics[width=0.7\textwidth]{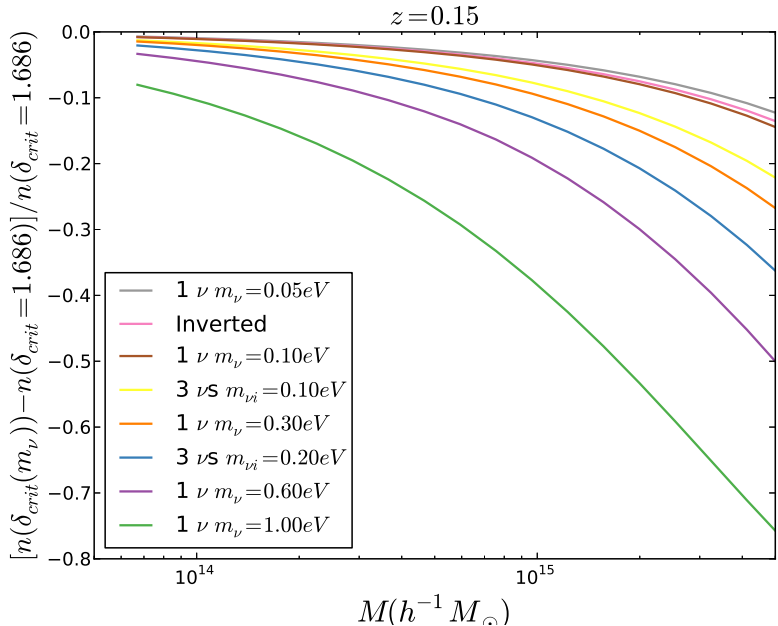} 
\caption{\label{fig:fracdeltanMs}  The fractional difference between the halo abundance $n(M)$ using $\delta_{crit} = 1.686$ and $n(M)$ calculated using our corrections to $\delta_{crit}(m_{\nu})$. Massive neutrinos increase $\delta_{crit}$ relative to $1.686$ so the abundance calculated assuming $\delta_{crit} = 1.686$ is larger. The order of the legend corresponds to the order of the curves.}
\end{center}
\end{figure}

In Fig.~\ref{fig:fracdeltanMs} we show the fractional difference between the mass function using our neutrino mass-dependent values of $\delta_{crit}$ and the standard value of $1.686$. The standard collapse threshold $\delta_{crit} = 1.686$ predicts more massive halos than our calculated $\delta_{crit}(m_{\nu})$. That is, our prescription for the halo mass function predicts that massive neutrinos cause greater changes to $n(M)$ than the standard prescription which leaves $\delta_{crit}$ fixed to the $\Omega_m = 1$ value. The fractional difference between our $n(M)$ and the standard prescription using $\delta_{crit} = 1.686$ increases slightly with increasing redshift. For $\sum_i m_{\nu ,i}\lsim 0.3eV$, the error in using $\delta_{crit} = 1.686$ remains $\lsim 10\%$  until $M= 10^{15}h^{-1} M_\odot$ for all redshifts between $z \sim 0 -1$.

\begin{figure}[t]
\begin{center}
\includegraphics[width=0.7\textwidth]{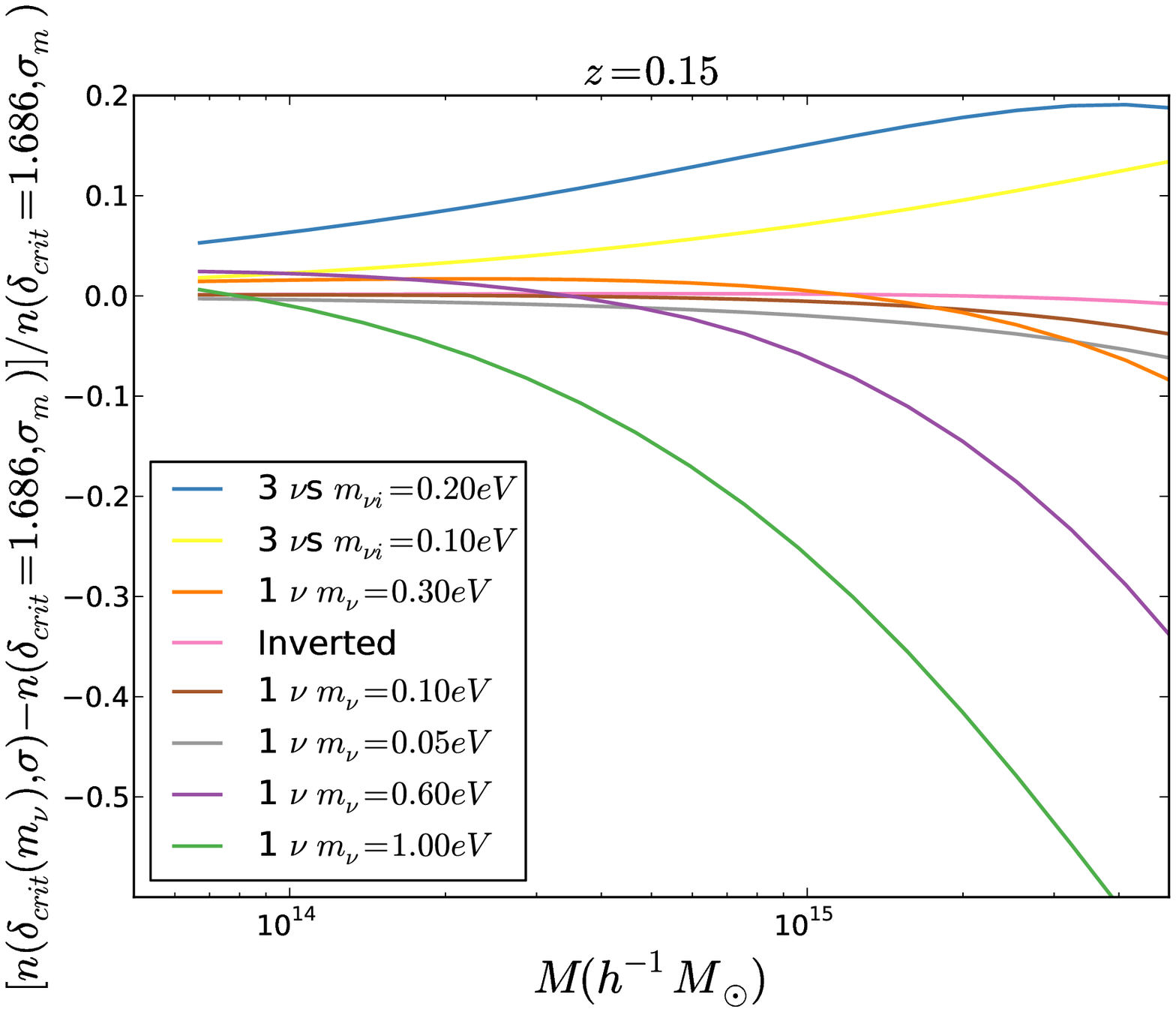}
\caption{\label{fig:fracdeltanMs_sigmaM} The fractional difference between the halo abundance $n(M)$ calculated using $\delta_{crit} = 1.686$ and $\sigma_m^2(M,m_{\nu})$ given in Eq.~(\ref{eq:sigma2mM}) (the variance of the total matter fluctuations including CDM, baryons, and neutrinos) -- as is sometimes done in the literature --  compared to our prescription using $\delta_{crit}(m_\nu)$ and the variance of CDM and baryons only ($\sigma^2(M)$ as in Eq.~(\ref{eq:sigma2M})). The order of the legend corresponds to the order of the curves.}
\end{center}
\end{figure}

Throughout this paper we have used the variance of CDM and baryon fluctuations ($\sigma(M)$  defined in Eq~(\ref{eq:sigma2M})) in the spherical collapse model (this is also the prescription of \cite{Ichiki:2011ue,Costanzi:2013bha}). A number of authors \cite{Brandbyge:2010ge,Marulli:2011he}, including the analyses of \cite{Mantz:2009rj, Benson:2011uta,Reichardt:2012yj}, have used the variance of the total (CDM + baryon + neutrino) mass fluctuations to calculate the mass function.  That is, in the halo mass function they have used $\sigma_m(M)$ 
\be
\label{eq:sigma2mM}
\sigma_m(M) \equiv \int \frac{dk}{k}\left|W(kR)\right|^2 \frac{k^3 P_{m}(k,z)}{2\pi^2}
\ee 
where $P_m$ is the total matter power spectrum including massive neutrinos, rather than $\sigma(M)$ calculated from the CDM and baryons alone (defined in Eq.~(\ref{eq:sigma2M})). In Fig.~\ref{fig:fracdeltanMs_sigmaM} we compare the difference between the halo mass functions calculated with our $\delta_{crit}$ from Eq.~(\ref{eq:deltacritsub}) and $\sigma(M)$ with the mass function calculated with $\delta_{crit} = 1.686$ and $\sigma_m(M)$. On halo scales massive neutrinos suppress $\sigma_m(M)$ more than $\sigma(M)$ so the combination of using $\delta_{crit} = 1.686$, which is lower than $\delta_{crit}(m_{\nu})$ and $\sigma_m(M)$ which is also lower than $\sigma(M)$ yields a final mass function that is closer to our prediction than the $\delta_{crit} = 1.686$, $\sigma(M)$ shown in Fig.~\ref{fig:fracdeltanMs}. 

\section{Conclusions}
\label{sec:conclusions}
In this paper we have extended the spherical collapse model to cosmologies with massive neutrinos with $m_{\nu }$ up to $1eV$. In our calculations of spherical collapse we take the effects of massive neutrinos into account in three ways: (i) in setting up the initial conditions for $R(t)$, (ii) in the background cosmology entering into the equation of motion for $R(t)$, and (iii) in that we allow for neutrino clustering in the halo during collapse. As expected, massive neutrinos delay the time of collapse ($R\rightarrow 0$). Including nonlinear clustering of neutrinos interior to the halo slightly decreases the delay in the collapse time but the net change due to massive neutrinos is still to delay collapse. 

In cosmologies with massive neutrinos the evolution of density perturbations becomes scale dependent once the neutrinos become non-relativistic. This scale-dependent evolution introduces some subtleties in mapping between spherical collapse solutions and halo abundance (see also \cite{Parfrey:2010uy} and \cite{Hui:2007zh, LoVerde:2014pxa} for a discussion of scale-dependent evolution and halo bias). We have developed an approach for studying spherical collapse in non-standard cosmologies with scale-dependent growth at early times which is outlined in \S \ref{sec:nuLCDM} and \S \ref{sec:massfcn}. We applied this to calculate the effect of massive neutrinos on the linearly extrapolated collapse threshold $\delta_{crit}(z_{collapse})$. We find that for cosmologies with $\sum_i m_{\nu i} \lsim 0.5eV$, the changes to $\delta_{crit}$ are $\lsim 1\%$, in agreement with previous works \cite{Brandbyge:2010ge,Ichiki:2011ue, Marulli:2011he, VillaescusaNavarro:2012ag, Costanzi:2013bha}. 

For scenarios with larger neutrino masses the changes to $\delta_{crit}$ are considerably larger. The dominant effects of neutrino mass on $\delta_{crit}$ come from two competing effects: the suppression in the growth of density perturbations on halos scales increases the collapse threshold, while non-linear clustering of massive neutrinos interior to the halo decreases the collapse threshold. The suppression in the growth of halo-scale perturbations dominates over the neutrino clustering in all the examples we considered. Interestingly, we found that the effects of massive neutrinos on $\delta_{crit}$ (and $\sigma(M)$) are not entirely characterized by $\Omega_\nu$. That is, we find that the predicted changes to $\delta_{crit}$ and $n(M)$ for cosmologies with different neutrino mass hierarchies (values of $m_{\nu 1}$, $m_{\nu 2}$, $m_{\nu 3}$) but the same $\Omega_\nu$ are not exactly the same (see Fig.~\ref{fig:delta_clustering_mndep}). This difference was also seen in the mixed dark matter simulations of \cite{Costanzi:2013bha}. In our framework the different dependence on $m_{\nu i}$ and $\Omega_\nu$ is predominantly due to the fact that the neutrino-induced suppression in the growth of structure depends on both the energy density in neutrinos and the redshift at which they became nonrelativistic (see e.g. \cite{Hu:1997mj,Lesgourgues:2006nd}). Clustering of neutrinos interior to the halo during collapse also causes changes to $\delta_{crit}$ that depend nonlinearly on the neutrino masses (Fig.~\ref{fig:deltac}) but the sense of this effect on $\delta_{crit}$ is opposite to that of the scale-dependent growth and is subdominant in all cases we considered.

In practice, distinguishing between neutrino mass hierarchies with halo abundance should be extraordinarily challenging. At the level of the spherical collapse model, the theoretical uncertainty in $n(M)$ is easily as large as the difference between the effects of different mass hierarchies (see e.g.  \cite{Bhattacharya:2010wy}). It is nevertheless important to note that the halo mass function should not be entirely determined by $\Omega_\nu$. We view the theoretical uncertainties here as additional motivation for N-body simulations with a range of neutrino mass hierarchies. 

An increasing number of authors have studied halo abundance in N-body simulations with mixed dark matter ($\nu \Lambda CDM$) cosmologies \cite{Brandbyge:2010ge,  Marulli:2011he, VillaescusaNavarro:2012ag,Costanzi:2013bha}. This literature includes simulations with three degenerate neutrino masses ranging from $m_{\nu i} = 0.05eV$ -- $ 0.40eV$ and measurements of halo abundance for masses  $M\lsim h^{-1} 10^{15}M_\odot$. For three degenerate neutrinos with $m_{\nu i} \lsim  0.2eV$ our predicted changes to $\delta_{crit}$ are $\lsim 1\%$ leading to a fractional change in $n(M)$ of only a few percent at $M = 10^{14}M_\odot$ but they give a suppression of $\sim 10\%$ at $M = 10^{15} M_\odot$ at $z = 0$ and $\sim 20\%$ at $z = 1$. Our predictions for the changes to $n(M,z)$ appear to be consistent with the halo abundance measured from simulations shown in Figure 2 in \cite{Costanzi:2013bha}. 

On the other hand, for $\sum_i m_{\nu i} = 1.2eV$, we predict that $\delta_{crit}$ is $\sim 2\%$ larger at $M = 10^{15} M_\odot$ which leads to a suppression in $n(M)$ at this mass of $\sim 30\%$ for fixed $\sigma(M)$. The authors of \cite{Brandbyge:2010ge} found that the Sheth-Tormen mass function with $\delta_{crit} = 1.686$ and $\sigma(M)$ calculated using the total matter power spectrum describes the halo abundance at $z=0$ in their simulations to better than $\sim 5 -10\%$. The difference between $n(M)$ predicted here (calculated with the larger value of $\delta_{crit}$ and $\sigma(M)$ calculated using the CDM + baryon power spectrum) and the approach of \cite{Brandbyge:2010ge} (using $\delta_{crit} = 1.686$ and $\sigma_m(M)$ calculated using the total matter power spectrum) is $10- 15\%$ for $\sum_i m_{\nu i} = 1.2eV$ at $z \sim 0$ and larger at higher redshift, which is marginally inconsistent with  \cite{Brandbyge:2010ge}.  It would be interesting to study the halo mass function in N-body simulations for the more extreme scenarios considered here (e.g. $m_{\nu i }\sim 1eV$) where we find more significant corrections to $\delta_{crit}$ and $n(M)$ (e.g. Fig. \ref{fig:deltanMs}). The difference between different prescriptions for the mass function is small for $\sum_i m_{\nu i}\lsim 0.3eV$, but for larger neutrino masses the variation between predictions is larger and could therefore alter the constraints on heavy neutrino species from cluster abundance.

In our discussion of the halo mass function we have neglected the neutrino contribution to the mass of the halos. For neutrino masses $\lsim 0.2eV$ the neutrino contribution to the halo mass should be $\lsim 10^{-3}$  so neglecting the neutrinos is justified \cite{LoVerde:2013lta}. On the other hand we found in \cite{LoVerde:2013lta} that for neutrino masses $\mathcal{O}(1eV)$ the neutrino mass associated with the halo could reach $\sim 10\%$ (though the neutrino mass interior to the virial radius is $\lsim 1\%$ of the CDM mass). Even a few percent shift in the halo masses can have a substantial effect on $n(M)$ so we caution that Figs. \ref{fig:deltanMs},  \ref{fig:fracdeltanMs}, and \ref{fig:fracdeltanMs_sigmaM}, which are written in terms of the CDM mass only, do not directly give the corrections to the observed halo abundance. The density profile of neutrinos is different from the density profile of CDM \cite{Kofman:1995ds,VillaescusaNavarro:2012ag, LoVerde:2013lta} so it is not obvious how the neutrino contribution will change the inferred halo mass and the answer will likely depend on the observable. We leave this question to further study.

\acknowledgements
M.L. thanks Brad Benson, George Fuller, Daniel Grin, Wayne Hu, Doug Rudd, and especially Matias Zaldarriaga for helpful discussions. M.L. is grateful for hospitality at the Institute for Advanced Study while this work was being completed. M.L. is supported by U.S. Dept. of Energy contract DE-FG02-13ER41958.

\appendix

\section{Initial conditions for $\delta_{cb}(t)$}
\label{sec:ICsfromCAMB}
The CAMB code gives the values of $\delta_c(k, z)$, $\delta_b(k, z)$, and $\delta_\nu(k, z)$ for a unit-magnitude initial curvature perturbation in the adiabatic mode. We modify the CAMB code so that the time derivatives of the density perturbations in CDM and baryons, $\dot{\delta}_c(z,k)$ and $\dot{\delta}_b(z,k)$,  are output as well.  To set up the initial conditions for $R$ and $\dot{R}$ we need $d\ln \delta_{cb}/dt(k) = \dot{\delta}_{cb}(k,z )/\delta_{cb}(k,z)$ (which is independent of $\delta_{cb}$ for linear perturbations).

\begin{figure}[t]
\begin{center}
$\begin{array}{cc}
 \includegraphics[width=0.5\textwidth]{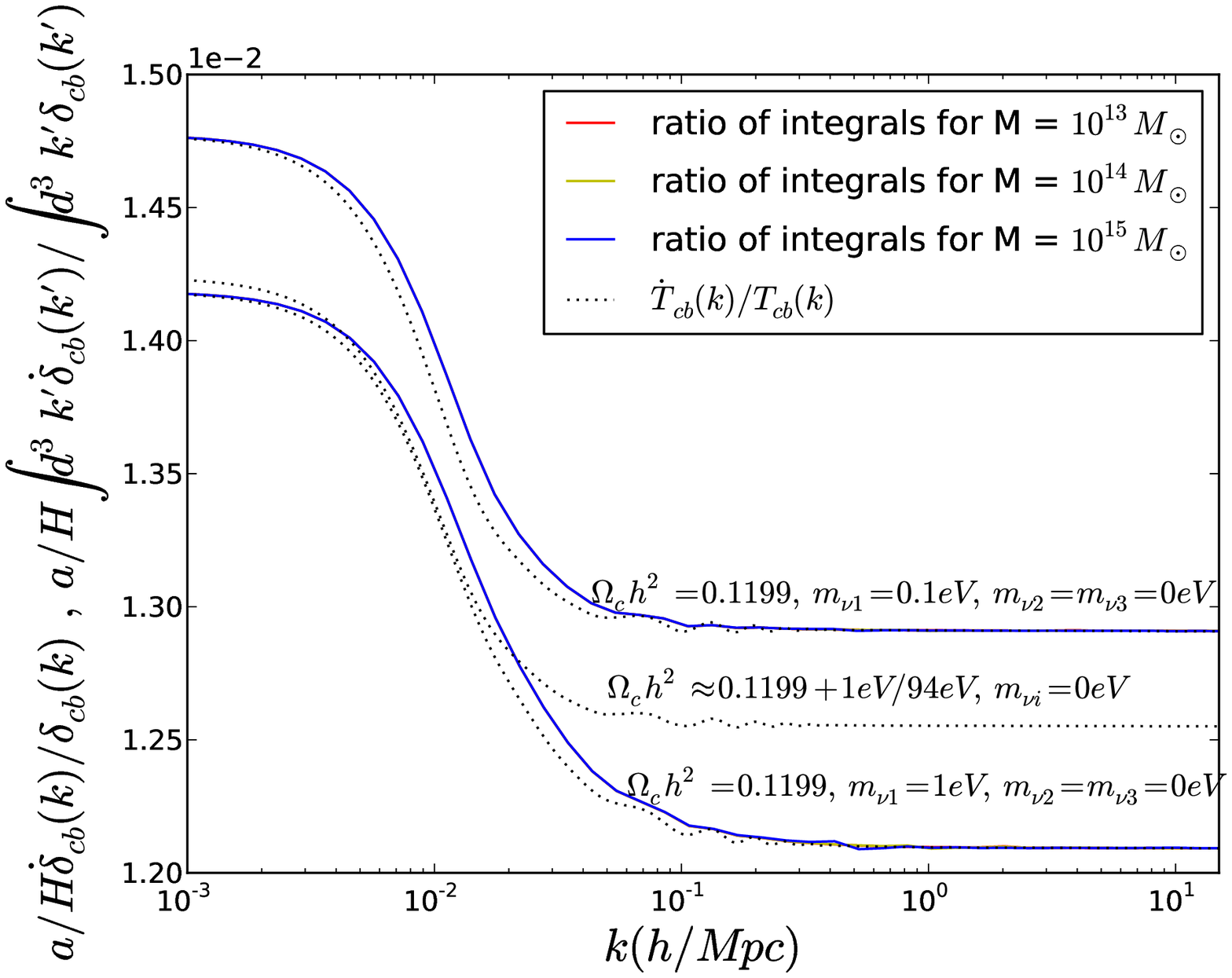} & \includegraphics[width=0.5\textwidth]{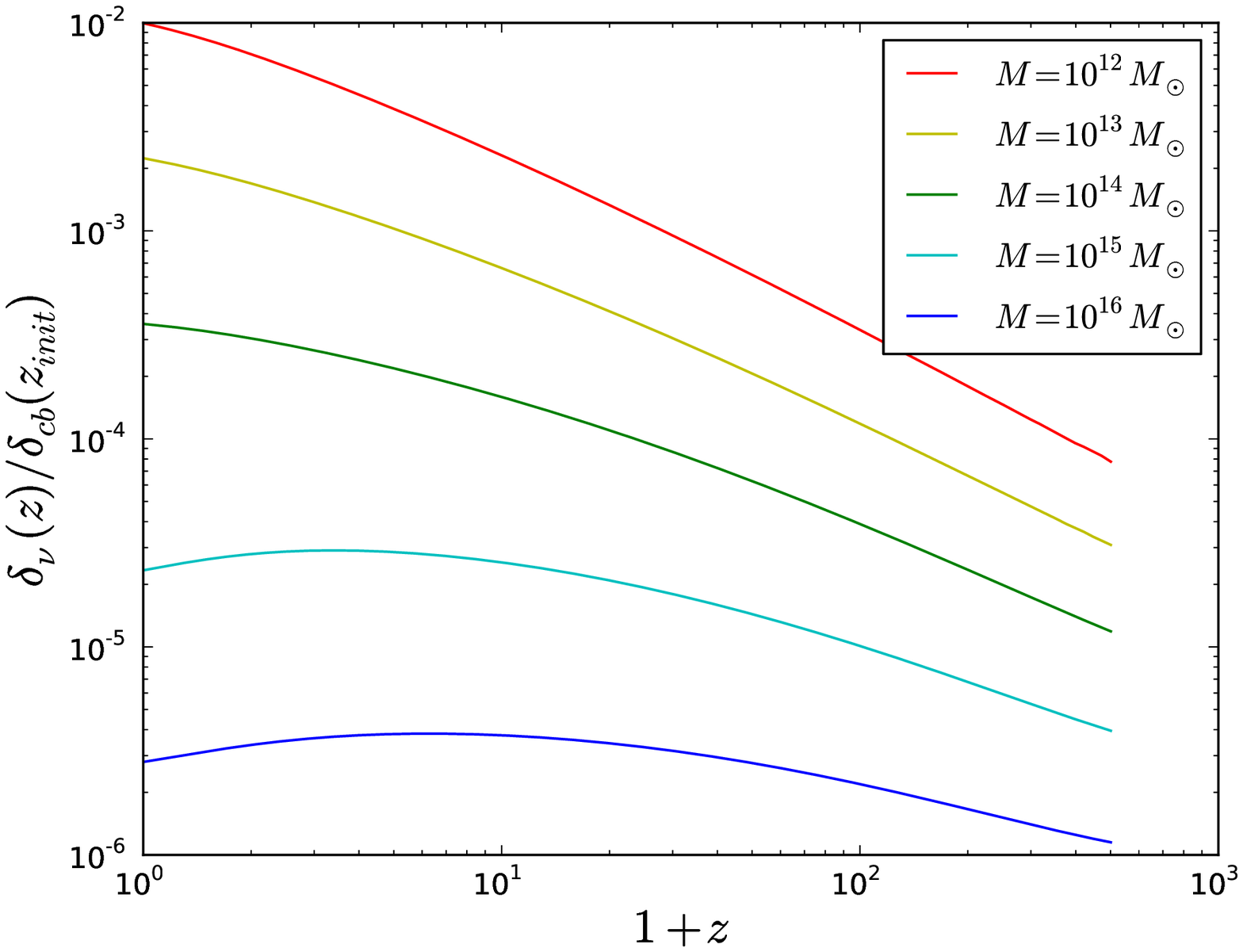}\\
  \mbox{(a)} &  \mbox{(b)} 
\end{array}$
\caption{\label{fig:deltaichecks} Left: The initial $\dot\delta_{cb}(z_{init})$ from CAMB. The solid curves show Eq.~(\ref{eq:deltadotics}) as a function of $k$, the curves for $M = 10^{13} M_\odot,\, 10^{14} M_{\odot}$, and $10^{15} M_\odot$ are indistinguishable. The dotted curves are the ratio $\dot{\delta}_{cb}(k,z_{init})/\delta_{cb}(k,z_{init})$ -- which at high $k$ is nearly identical to Eq.~(\ref{eq:deltadotics}). All curves are at $z = 200$ in units of $H(z)$.   Right: The linear neutrino perturbation, $\delta_\nu(z) = \int d^3 {\k}\, {T_\nu(k,z)}/{T_{cb}(k,z_{init})}\delta_{cb}(z_{init}) W(kR)$,  plotted as a function of redshift in units of $\delta_{cb}(z_{init})$ for a range of halo masses.}
\end{center}
\end{figure}

For an initial density perturbation with a top-hat profile, the initial velocity is given by
\be
\label{eq:dlndelta}
\frac{d\ln \delta_{cb}}{dt} = \frac{\int d^3 \k\, W(kR) \dot\delta_{cb}(k,z_{init})/\delta_{cb}(k,z)}{\int d^3 \k\, W(kR) }\,,
\ee
where $W(kR) =   3j_1(kR(M)/(kR(M))$ and $\dot\delta_{cb}(k,z)/\delta_{cb}(k,z)$ is the ratio of the amplitudes in each Fourier mode which can be obtained from the transfer functions given by CAMB. Note that for $\delta_{cb}$ independent of $k$, the integrals just return a constant. That is, 
\be
\label{eq:deltafcn}
\int \frac{d^3 k}{(2\pi)^3} W(kR) = \int d^3k  \int_{\frac{4}{3}\pi R^3} d^3x e^{- i k\cdot x} =  1\,,
\ee
but Eq.~(\ref{eq:deltafcn}) doesn't converge numerically. However, the ratio of the two expressions in Eq.~(\ref{eq:dlndelta}) plotted as a function of the maximum value $k$ does reach an asymptote. In Figure \ref{fig:deltaichecks} we plot 
\be
\label{eq:deltadotics}
\frac{d\ln \delta_{cb}}{dt}(k) = \frac{\int_0^{k} k'^2 dk'\, W(k'R) \dot\delta_{cb}(k',z_{init})/ \delta_{cb}(k',z_{init})}{\int_0^{k} k'^2d k'\, W(k'R)}\,,
\ee
along with the ratio $\dot\delta_{cb}(k,z)/\delta_{cb}(k)$. In Fig. \ref{fig:deltaichecks} we see that $\dot\delta_{cb}(z_{init})$ as given in Eq.~(\ref{eq:deltadotics}) does not depend on halo mass for the range of masses considered but it does depend on the neutrino masses -- the difference at low $k$ is due to the different $\Omega_m(z_{init})$'s but the larger suppression at high $k$ is due to $m_{\nu} >0$. We can also see that by $k\sim 1 h/Mpc$ the values of Eq.~(\ref{eq:deltadotics}) and $\dot\delta_{cb}(k,z)/\delta_{cb}(k)$ nearly identical. We therefore use $d\ln \delta_{cb}/dt = \dot\delta_{cb}(k,z)/\delta_{cb}(k)$ evaluated at $k= 10Mpc^{-1}$ to determine the initial conditions of $R$ for a top-hat initial density perturbation. As discussed in \S \ref{sec:massfcn} the calculations of the critical overdensity in the plots of the body of this paper use the root-mean-square value of the initial velocity given in Eq.~(\ref{eq:dotsigma2M}), rather than Eq.~(\ref{eq:deltadotics}). We have used the top-hat initial conditions of Eq.~(\ref{eq:deltadotics}) in Fig.~\ref{fig:Roft}, Fig.~\ref{fig:comparedelta1000}, and Fig.~\ref{fig:deltacTH}.

For adiabatic initial conditions there are perturbations in the other components $\delta_\nu$ and $\delta_\gamma$ at the initial time as well. A top-hat initial density perturbation in CDM and baryons with amplitude $\delta_{cb,init}$  and radius $R$ has Fourier components given by 
\be
\delta_{cb} (\k)= \delta_{cb,init} W(kR)\,,
\ee
which allows us to determine the amplitude of the primordial curvature perturbation in each Fourier mode, and therefore the perturbations in the other components
\be
\delta_{cb,i}(\k,z_{init}) = T_{cb}(k,z_{init}) \zeta(\k)  \rightarrow \delta_\nu(\k, z_{init}) = \frac{T_\nu(k,z_{init})}{T_{cb,i}(k,z_{init})}\delta_{cb}(z_{init}) W(kR)\,,
\ee
where $\zeta(\k)$ is the primordial curvature. The linear evolution of the initial density perturbation, $\delta_\nu(z) = \int d^3 {\k}\, {T_\nu(k,z)}/{T_{cb}(k,z_{init})}\delta_{cb}(z_{init}) W(kR)$, for a top-hat CDM and baryon perturbation is plotted in Figure \ref{fig:deltaichecks}. 

At our start time, $z_{init} = 200$, the amplitude of density perturbations that collapse by $z\lsim 1$ is $\sim 10^{-2}$. The amplitude of non-linear corrections to our initial conditions (which we have ignored) is $\mathcal{O}(\delta_{init}^2) \sim 10^{-3}-10^{-4}$ which is not very different from the magnitude of the neutrino mass effects on $\delta_{crit}$ for some of the smaller neutrino masses. However, to make non-linear corrections completely subdominant we would need to start our calculation before decoupling and then follow the subhorizon evolution of the baryons independently of the CDM until $z\sim 200$. The total density perturbation would no longer have a top-hat profile in this case and this seems like overkill for a model that is anyway intended to be a crude approximation to halo formation. However, we have tested the sensitivity of our calculations to the initial conditions in two ways: (i) by changing $z_{init}$ to earlier times and repeating the spherical collapse calculation, and (ii) by shifting $z_{init}$ but keeping $\delta_{init}$ fixed between calculations with different values of the neutrino mass (so that the magnitude of the non-linear corrections are similar). In both cases the changes to $(\delta_{crit}(m_{\nu}) - \delta_{crit}(m_\nu= 0))/\delta_{crit}$ are $\lsim (few) 10^{-3}$. Changing the ratios of CDM to baryons has an effect that is comparable in magnitude. We therefore consider our calculations of $\delta_{crit}$ to be accurate (independent of $z_{init}$ and $\Omega_b$) to about $\sim 0.5\%$.

\section{How accurate is the expression for $\delta M_\nu$? Does it matter for $\delta_{cb}^{linear}(z_{collapse})$?}
\label{sec:deltaMnutests}
In \S\ref{ssec:deltarhonu} we used a solution to the linearized Boltzmann equation to calculate the neutrino mass interior to $R$, $\delta M_\nu(<R)$. The linearized Boltzmann solution is accurate at early times before the CDM density fluctuation has become nonlinear, but significantly underestimates the late-time non-linear clustering of neutrinos (e.g. \cite{Ringwald:2004np, LoVerde:2013lta}). In this section we explore the sensitivity of our results to the approximation for $\delta M_\nu(<R)$ used in \S \ref{ssec:deltarhonu}. Our reference point for a more accurate calculation of $\delta M_\nu$ numerically integrates neutrino trajectories traveling in the external potential of the collapsing halo. The initial conditions for the neutrino trajectories sample the initial phase space and $\delta M_\nu(<R, t)$ is obtained by summing elements of phase space with trajectories interior to $R$ at time $t$ (for details see \cite{LoVerde:2013lta}). We refer to this calculation as the ``exact" calculation because it is an exact solution to the Boltzmann equation for neutrinos in an external potential. Our exact calculation, however,  still makes the approximation that $\delta M_\nu$ can be calculated using a halo potential described by the solution to $R(t)$ with $\delta M_\nu = 0$. 

\begin{figure}[t]
\begin{center}
$\begin{array}{cc}
 \includegraphics[width=0.5\textwidth]{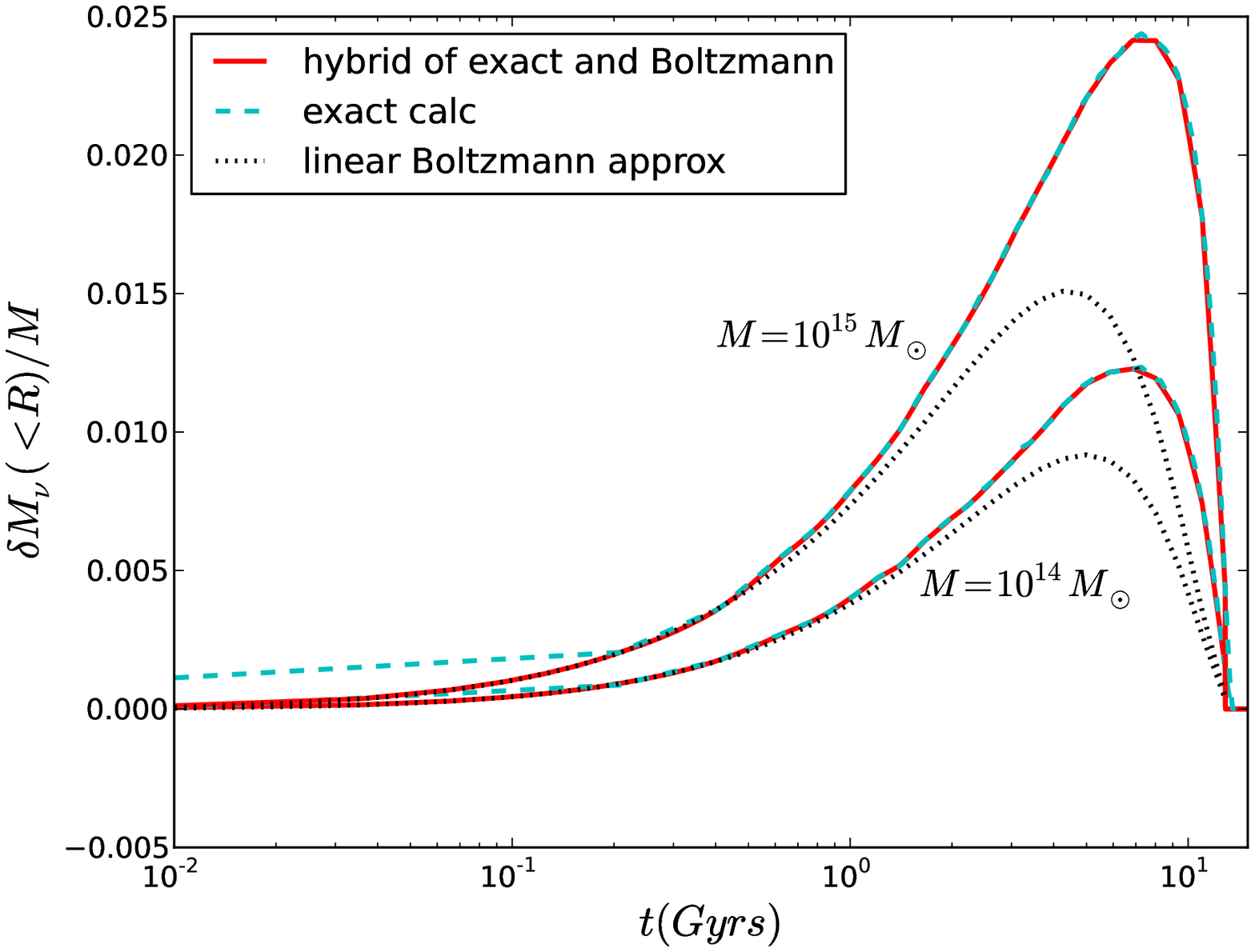} & \includegraphics[width=0.5\textwidth]{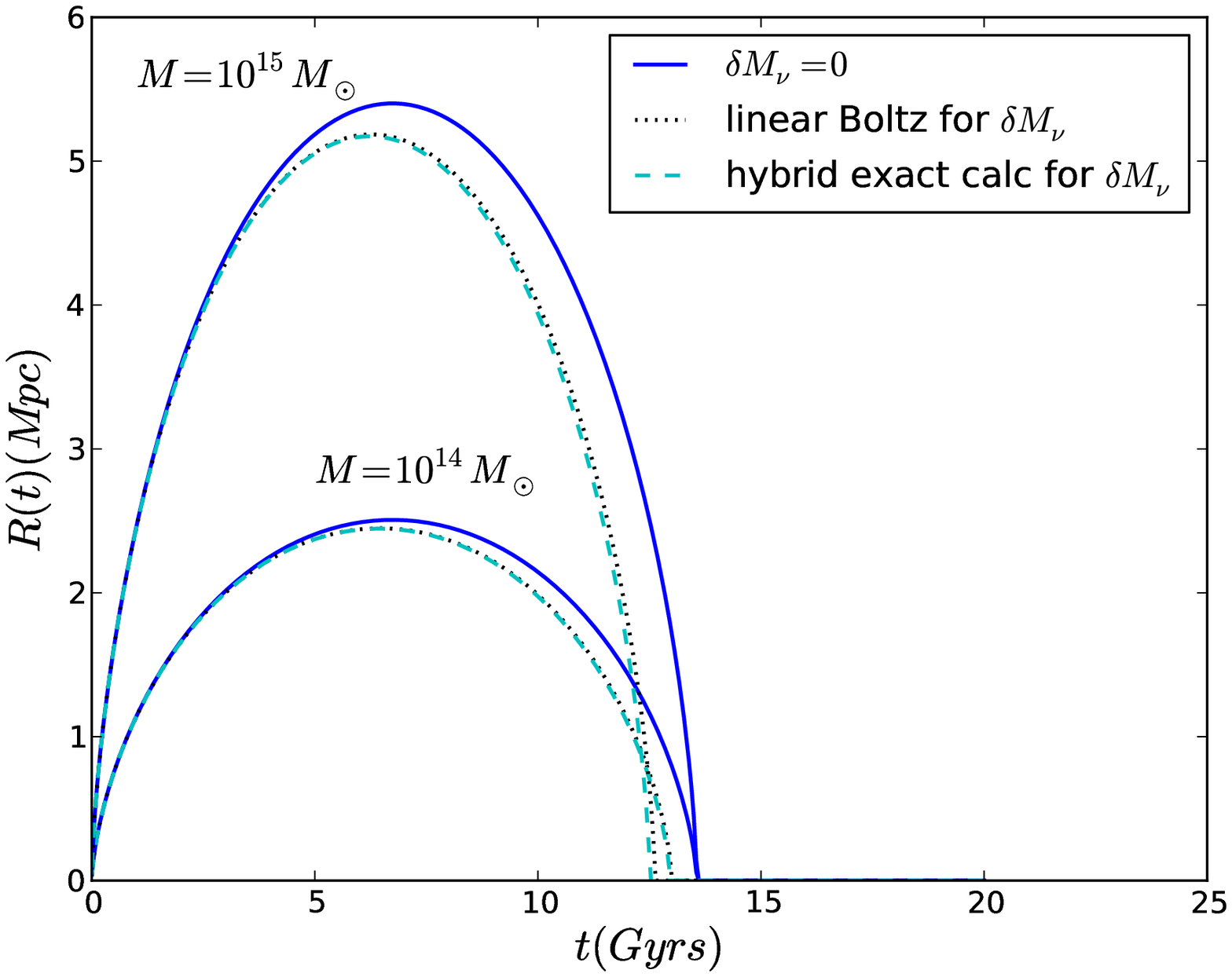}\\
  \mbox{(a)} &  \mbox{(b)} 
\end{array}$
\caption{\label{fig:deltaMnuchecks} Left: Several calculations of the neutrino mass interior to $R$, $\delta M_\nu$, plotted for a cosmology with one massive neutrino with $m_{\nu} = 1eV$ and halos of $M = 10^{14} M_\odot$ and $M = 10^{15} M_\odot$. Shown is the linearized Boltzmann solution, the exact calculation, and a hybrid solution that uses the linearized Boltzmann solution at early times, where the exact calculation hasn't converged and the exact calculation at late times once $\delta M_{\nu, {\rm exact}} >  \delta M_{\nu, {\rm linear}}$. Right: The solutions to Eq.~(\ref{eq:ddotR}) using $\delta M_\nu = 0$, the linearized Boltzmann solution for $\delta M_\nu$, and the hybrid exact-linear solution shown in the right panel. Despite the large differences in $\delta M_\nu$ in the right panel, the solutions for $R(t)$ are very similar -- the collapse time changes by $\lsim 1\%$ for $M = 10^{15} M_\odot$ and $< 0.5\%$ for $M = 10^{14} M_\odot$. These changes in $t_{collapse}$ and are small compared to the change due to the effect we are interested in, nonzero $\delta M_\nu$.}
\end{center}
\end{figure}

Figure (\ref{fig:deltaMnuchecks}) shows the difference between the linearized Boltzmann solution for $\delta M_\nu$ and the exact calculation for halos with $M = 10^{14} M_\odot$, $M = 10^{15} M_\odot$,  and single massive neutrino species with $m_\nu = 1eV$. The difference between the exact calculation and the linear approximation is large. Plotted in the right panel of Figure (\ref{fig:deltaMnuchecks}) are the solutions for $R(t)$ from Eq.~(\ref{eq:ddotR}) with $\delta M_\nu =0$, $\delta M_{\nu}$ from the linearized Boltzmann equation, and $\delta M_{\nu}$ using the linearized Boltzmann equation at early times and the exact calculation once $\delta M_{\nu, {\rm exact}} >  \delta M_{\nu, {\rm linear}}$. As discussed in \S \ref{sec:nuLCDM} non-zero $\delta M_\nu$ has a large effect on the evolution of $R(t)$ and the collapse time. However, the difference between $R(t)$ including the larger $\delta M_{\nu{\rm exact}}$ and the linearized $\delta M_{\nu{\rm linear}}$ is negligible. For the examples we have considered, halos with $M = 10^{14} M_\odot$ and $M = 10^{15} M_\odot$, the collapse times change by $< 0.5\%$, $< 1\%$. Apparently, the effect of $\delta M_\nu$ on the final collapse time of $R$ is most important at early times. Nonlinear clustering of neutrinos depends strongly on the neutrino mass. Since the error in $t_{collapse}$ from using the linearized Boltzmann calculation for $\delta M_\nu$ is unimportant for $m_{\nu} = 1eV$ we conclude that it is a safe approximation for all neutrino mass hierarchies considered in this paper. 

\section{Calculating $\delta_{crit}(z)$ in a cosmology with scale dependent evolution}
\label{sec:comparedeltacrit}

There are a number of analytic models of halo abundance, from the original Press-Schechter ansatz \cite{Press:1973iz} to more sophisticated excursion set and peaks calculations (\cite{Bond:1990iw,Bardeen:1985tr}).  These models make use of the fact that the linear evolution of the density field is well understood and identify regions of the early-time, linear density field that satisfy certain criteria, with halos at late times. For instance, a halo of mass $M$ forming at redshift $z$ can be associated with a region in the early-time density field $\delta_{cb,init}$ smoothed on scale $R = (3M/(4\pi\rho_{cb}))^{1/3}$ that exceeds the threshold for collapse ($\delta_{cb}(z_{init}) >\delta_{crit}(z_{init})$) and/or is a peak ( $\nabla \delta_{cb}(\x,z_{init}) = 0$ and $\det(\nabla_i\nabla_j \delta_{cb}(\x,z_{init})) < 0$).  If the statistics of the linear density field are known (in the standard cosmology they are Gaussian) then the fraction of the initial volume, and therefore the mass, that satisfies the halo criteria (peaks or thresholds) can be calculated. 

For instance, the excursion set model relates the abundance of halos at $z$ to the distribution of first crossings of the barrier $\delta_{crit,i}(z)$ which (in the limit of a Fourier-space top-hat smoothing function) gives the usual Press-Schechter mass function
\be
\label{eq:nPS}
n_{PS}(M,z) =  - 2\frac{\bar{\rho}_{cb}}{M} \frac{d}{dM}\left[\int_{\frac{\delta_{cb}(z_{collapse},z_{init})}{\sigma(M,z_{init})}}^\infty d\nu \frac{e^{-1/2\nu^2}}{\sqrt{2\pi}}\right]  \,,
\ee
where $\delta_{cb}(z_{collapse},z_{init})$ is the value of the density fluctuation at $z_{init}$ required to collapse by redshift $z_{collapse}$. Typically, one works with the collapse threshold and the density field linearly extrapolated to the present time. This is a convenient choice because $\delta_{crit}(z_{collapse},z_{collapse})$ is nearly independent of redshift. And, for a cosmology with scale-independent growth a top-hat density perturbation (or a density perturbation of any profile) $\delta_{cb}(z)$ evolves in the same way as $\sigma(M,z)$ so the ratio $\delta_{cb}(z)/\sigma(M,z)$ is constant. 

In cosmologies with massive neutrinos, or any scale-dependent growth, the growth rates of density perturbations depend on their profile. To be completely explicit, 
\be
\delta_{cb}(z) = \int \frac{d^3\k}{(2\pi)^3} \frac{T_{cb}(k,z)}{T_{cb}(k,z_{init})}\delta_{cb}(k,z_{init})
\ee
where $T_{cb}(k,z)$ are the transfer functions. Furthermore, the scale-dependent growth will change the density profiles of perturbations -- e.g. an initially top-hat density perturbation can evolve into a perturbation with a slightly different profile.  

Quantities that describe the statistics of the density field will evolve differently from individual density perturbations. For instance, the variance of density fluctuations evolves as 
\be
\sigma^2(R,z) = \int \frac{d^3\k}{(2\pi)^3}|W(kR)|^2 \left|\frac{T_{cb}(k,z)}{T_{cb}(k,z_{init})}\right|^2 P_{cb}(k,z_{init})
\ee
so the ratio $\delta_{cb}(z)/\sigma(M,z)$ may depend on redshift $z$ because the numerator and denominator are weighted towards different $k$ ranges. On the other hand the ratio $\delta_{cb}(k)/\sqrt{k^3 P_{cb}(k)}$ is time independent. However one has to be cautious in using the statistics of the initial density field smoothed on some scale $\delta_{cb, R}(z_{init})$ to define halos at late times so that the description of halo abundance does not depend on the initial time. 

The issue of scale-dependent growth and the definition of the collapse threshold was also discussed by \cite{Parfrey:2010uy} in the context of modified gravity theories. In the example of \cite{Parfrey:2010uy}, the growth is scale independent at early times so that $\delta_{cb,R}(z_{init})$ is sufficient to specify $\dot{\delta}_{cb,R}(z_{init})$. Those authors noted that the relevant quantity for calculating the halo abundance is $\delta_{cb,R}(z_{init})/\sigma(M,z_{init})$. To evaluate $\delta_{crit}$ at a different redshift while preserving the ratio $\delta_{cb,R}(z_{init})/\sigma(M,z_{init})$ one would use $\delta_{crit}(z) \equiv \delta_{cb,R}(z_{init})\sigma(M,z)/\sigma(M,z_{init})$ which is consistent with Eq.~(\ref{eq:deltacrit}) in this paper.  However, in the scenario of \cite{Parfrey:2010uy} the scale-dependent growth does not become important until late times so the initial velocity of a top-hat perturbation and the root-mean-square value of top-hat perturbations $\dot{\sigma}(M,z_{init})/\sigma(M,z_{init})$ are the same and there is no ambiguity in setting the initial conditions for $R(t)$. 

\begin{figure}
\begin{center}
$\begin{array}{cc}
\includegraphics[width=0.5\textwidth]{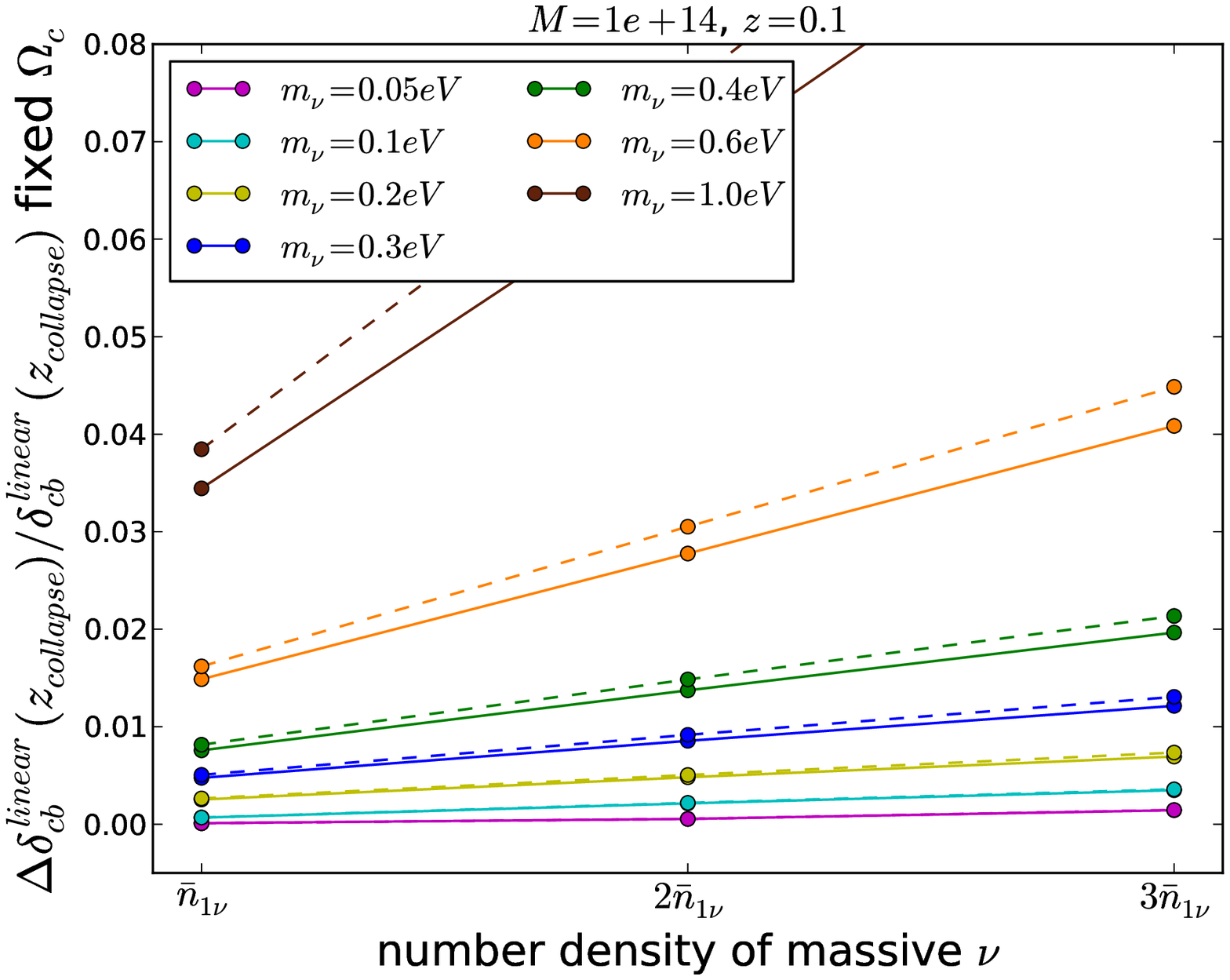} & \includegraphics[width=0.5\textwidth]{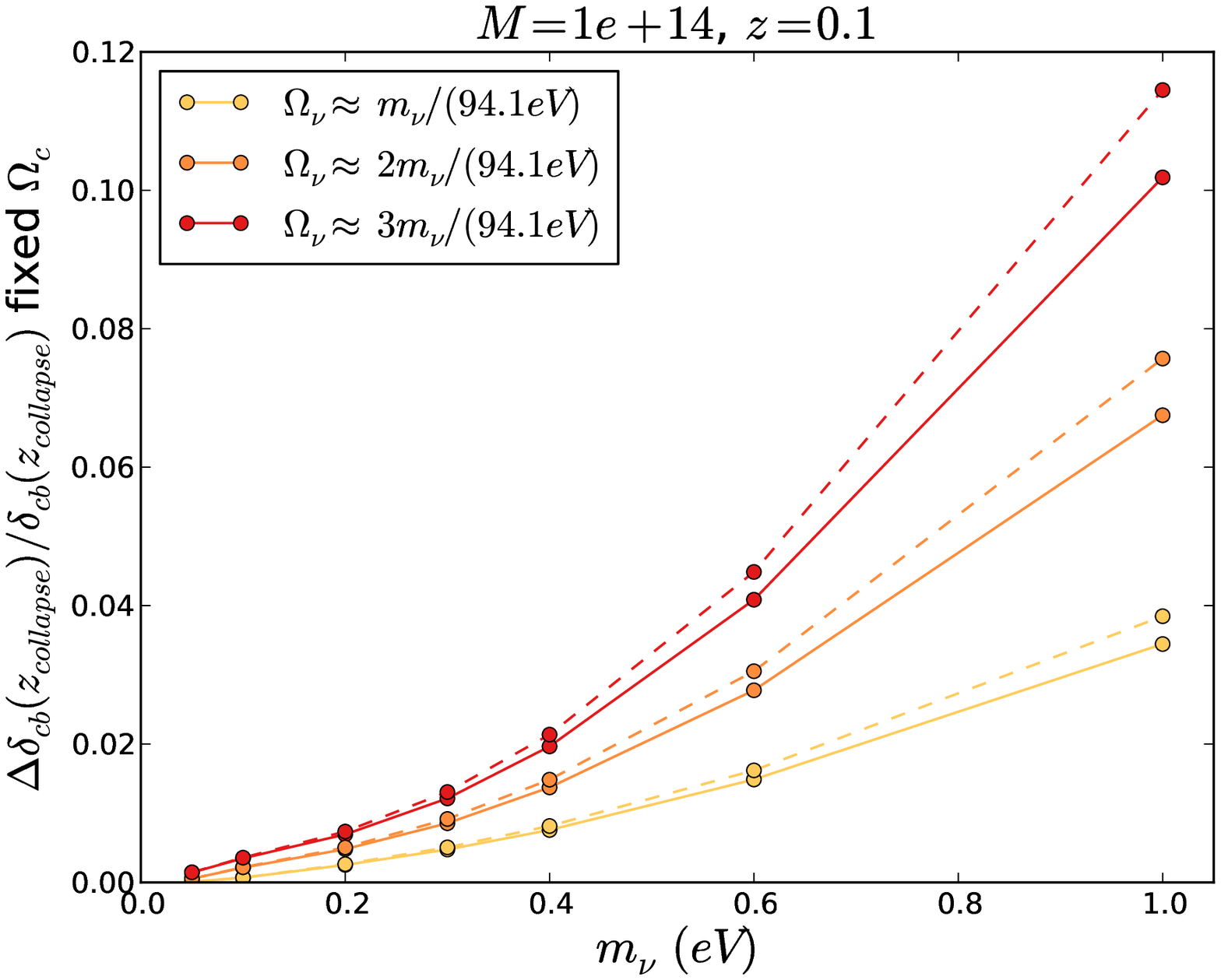}\\
\mbox{(a)} & \mbox{(b)}
\end{array}$
\caption{\label{fig:comparedelta1000} A comparison of the neutrino-induced changes to the linearly extrapolated collapse threshold calculated in two ways: first, using the initial velocity $\dot{\delta}_{cb}(z_{init})/\delta_{cb}(z_{init})  \equiv \dot{\sigma}/\sigma$ and the linear extrapolation given in Eq.~(\ref{eq:deltacrit}) as in the body of this paper (solid lines); second, using the initial conditions and linear extrapolation of a top-hat perturbation at $z = 1000$ (dashed lines). The two approaches are in agreement at the $\lsim 20\%$ level.}
\end{center}
\end{figure}
In this paper we have chosen to use $\dot{\sigma}(M,z_{init})/\sigma(M,z_{init})$ to set the initial conditions for $R(t)$ for two reasons: (i) this choice represents typical initial conditions for $\delta_{cb,R}(\x)$, and (ii) it allows Eq.~(\ref{eq:deltacrit}), which should represent the rarity of initial conditions that collapse by $z$, to be independent of $z_{init}$. Note that there still may be residual sensitivity to the initial time, due to the fact that we set initial conditions using $\sigma(M,z)$ which does not evolve according to the same equation of motion as the linear perturbation $\delta_{cb}$. In principle we could start our calculations at earlier times before neutrinos become nonrelativistic (hence before they induce scale-dependent growth) so that we would avoid the need to make this choice for $\dot{\delta}_{cb,R}$. Unfortunately this does not get around the issue because at such early times the linear evolution is scale dependent because of the presence of radiation and the fact that baryons are not exactly tracking CDM. However, as a sanity check we can compare our predicted change to the collapse threshold in the presence of massive neutrinos from Eq.~(\ref{eq:deltacrit}) and Eq.~(\ref{eq:sigmadeltadot}) to an alternative definition of $\delta_{crit}$ using top-hat initial conditions at early times when the scale-dependent growth due to massive neutrinos is small. Specifically, we define
\be
\label{eq:deltacrit1000}
\delta_{crit}^{{\rm {\tiny early}}}(z_{collapse}) = \frac{\delta^{{\rm {\tiny top-hat}}}_{crit}(z = 1000)}{\sigma(M,z = 1000)}\sigma(M,z_{collapse})
\ee 
where $\delta^{{\rm {\tiny top-hat}}}_{crit}(z = 1000)$ is the critical value of a top-hat density perturbation, with $\dot{\delta}_{cb}$ calculated for top-hat initial conditions as in Appendix \ref{sec:ICsfromCAMB} linearly extrapolated to $z = 1000$.  This definition is in the same spirit as the approach of \cite{Parfrey:2010uy}. A comparison between the neutrino corrections to $\delta_{crit}$ calculated using Eq.~(\ref{eq:deltacrit1000}) and Eq.~(\ref{eq:deltacrit}) is plotted in Fig.~\ref{fig:comparedelta1000}. The neutrino corrections from the two calculations are in agreement at the $\lsim 20\%$ level. 

Another approach for treating the scale-dependent growth would be to determine the correlated requirements on both $\delta_{cb,R}(z_{init})$ and $\dot{\delta}_{cb,R}(z_{init})$ for a halo to collapse by a given time. Then one could associate regions in the initial density field that satisfy the joint criteria with halos forming at the collapse redshift. Developing such a model would be quite interesting but is beyond the scope of this paper. 

\begin{figure}[t]
\begin{center}
$\begin{array}{cc}
 \includegraphics[width=0.5\textwidth]{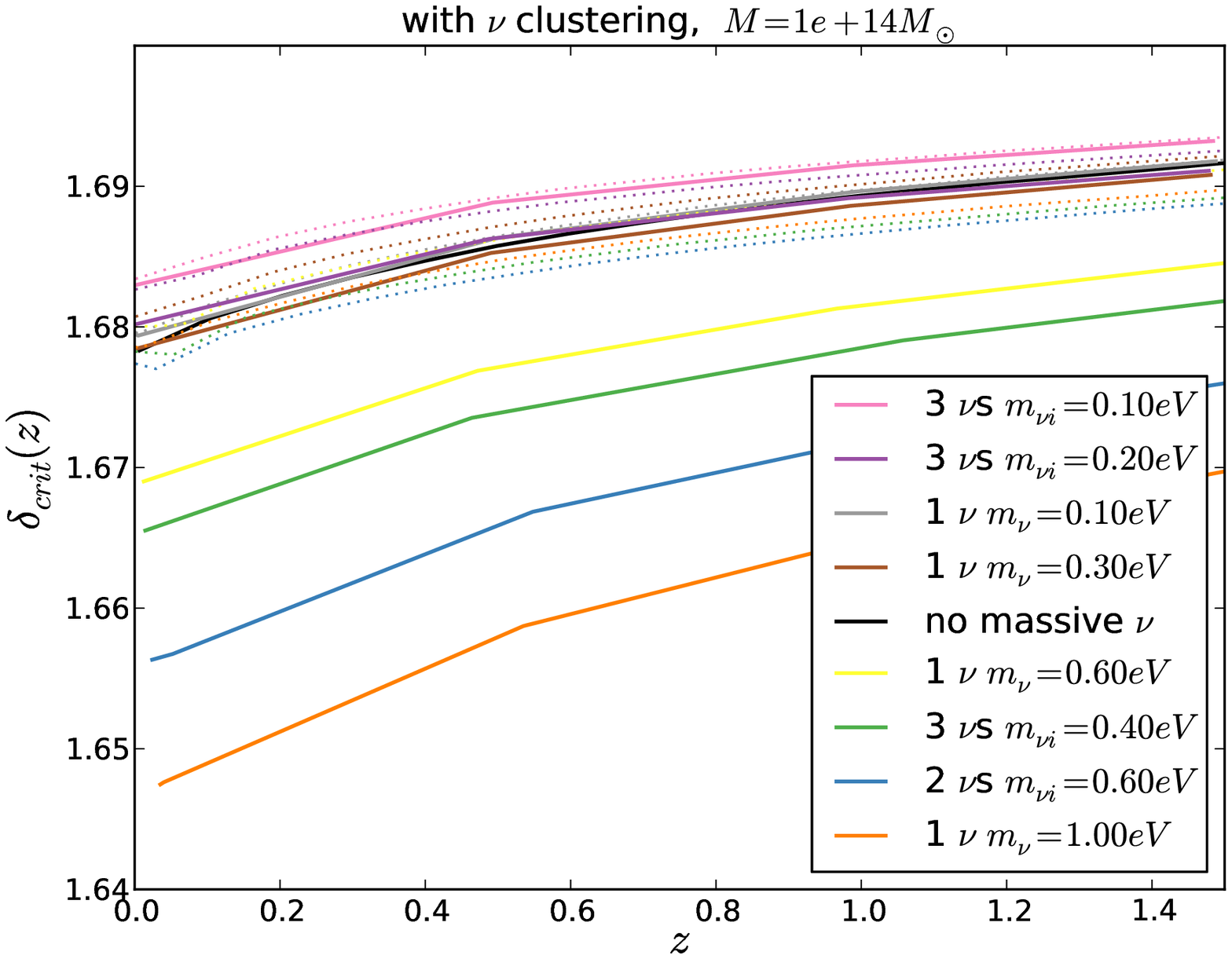}&   \includegraphics[width=0.5\textwidth]{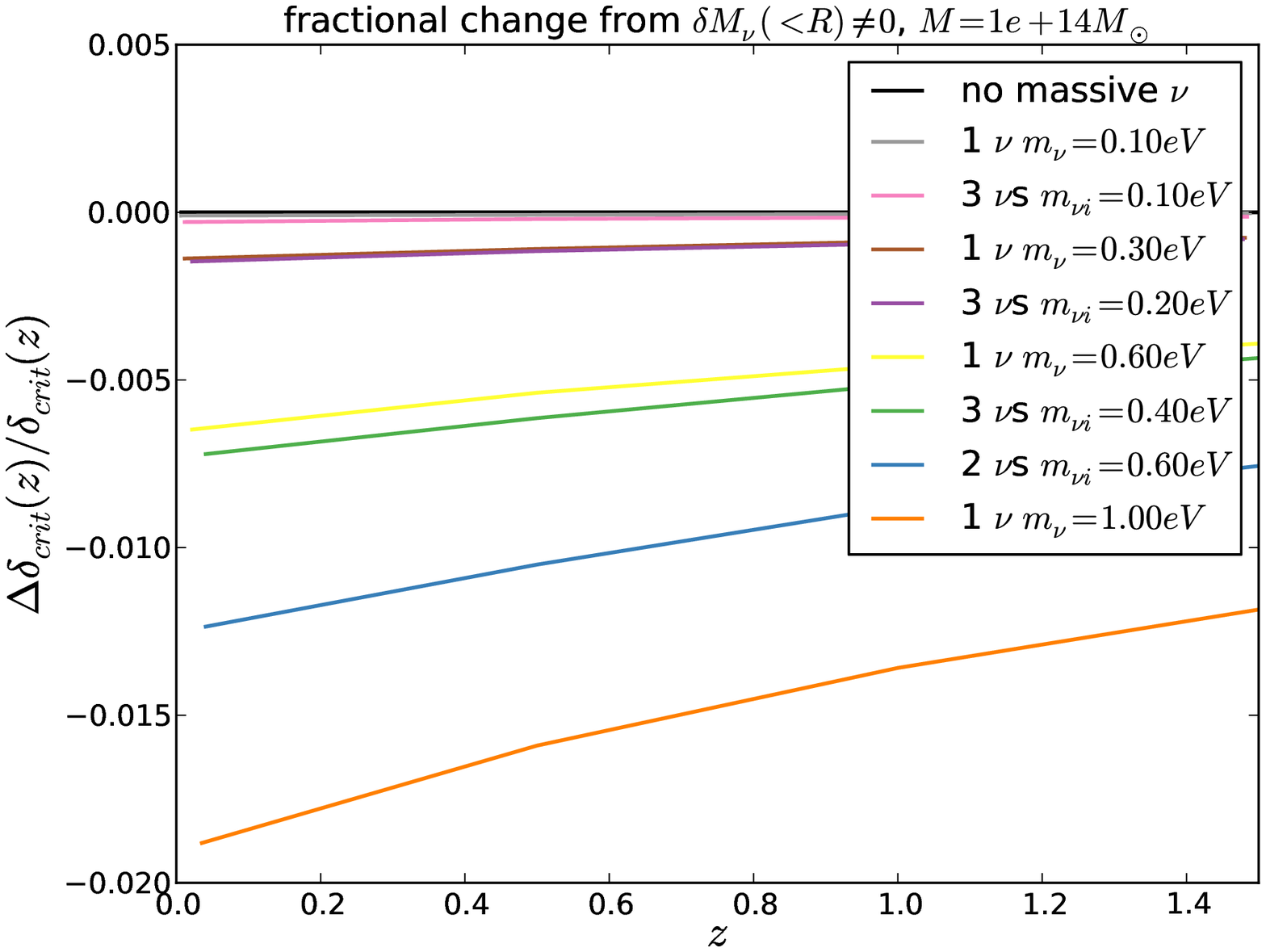} \\
 \mbox{(a)} & \mbox{(b)}
\end{array}$
\caption{\label{fig:deltacTH} Left: The linearly extrapolated value of the initial amplitude of the density perturbation $\delta_{cb}(z_{init})$ required to collapse at $z_{collapse}$, where the initial $\dot{\delta}_{cb}(z_{init})$ and the linear extrapolation are done for an exactly top-hat density profile at $z = 200$ (see Eq.~(\ref{eq:deltaTHdef}) and surrounding text). The solid lines include neutrino clustering, while the fainter dotted lines neglect it ($\delta M_\nu(<R) =0$ in Eq.~(\ref{eq:ddotR})). Here we have fixed $\Omega_c$ and $\Omega_b$ so curves with different $\Omega_\nu$ have different total matter density $\Omega_m$. Right: The fractional change to the collapse threshold when neutrino clustering interior to $R$ is included. In both panels the order of the curves matches the order of the legends. }
\end{center}
\end{figure}

Finally, we have so far treated the threshold for collapse as a criteria on the initial density field, $\delta_{cb}(z_{init})$ with statistics characterized by the initial power spectrum $P_{cb}(k,z_{nit})$ and looked for consistent ways of imposing this criteria on the density field linearly extrapolated to late times. One could instead treat the linearly extrapolated value of of a top-hat perturbation $\delta_{cb}(z_{init})$ as the parameter of interest, for instance as a quantity in a fitting function.  That is, one could use the initial conditions of a true top-hat density perturbation $\delta_{cb,init}$, with $\dot{\delta}_{cb, init}$ as given in Eq.~(\ref{eq:deltadotics}), for the initial conditions for $R(t)$ and linearly evolve $\delta_{cb,init}$ to $z_{collapse}$ using the true top-hat evolution. This gives
\be
\label{eq:deltaTHdef}
\delta^{{\rm {\tiny top-hat}}}_{crit}(z) \equiv \frac{\int d^3k\, T_{cb}(k, z)/T_{cb}(k, z_{init}) W(kR)}{\int d^3k W(kR)}\delta_{cb,i}\,. %\approx \frac{T_{cb}(k*,z)}{T_{cb}(k*,z)}\delta_{cb}(z_{init}) \,.
\ee
This approach results in quite different values of $\delta_{crit}(z)$ for cosmologies with large scale-dependent growth near the halo scale (for massive neutrinos this happens when $m_{\nu,i} \gsim 0.3eV$). More importantly, in these cases $\delta_{crit}(z)/\sigma(M,z) \neq \delta_{cb,i}/\sigma(M,z_{init})$ so the linearly extrapolated $\delta^{{\rm {\tiny top-hat}}}_{crit}$ can not be straightforwardly interpreted as a collapse threshold. For comparison, we show these calculations in Fig.~\ref{fig:deltacTH}. Interestingly, the sensitivity to neutrino mass in $\delta^{{\rm {\tiny top-hat}}}_{crit}(z)$ is almost entirely due to the clustering of neutrinos interior to $R$. If $\delta M_\nu(<R) =0$, the ``top-hat" collapse threshold for cosmologies with massive neutrinos in Eq.~(\ref{eq:deltaTHdef}) is $\lsim 0.5\%$ different from Eq.~(\ref{eq:deltaTHdef}) for cosmologies with $m_{\nu i} =0$. The change to $\delta^{{\rm {\tiny}top-hat}}_{crit}$ from $\delta M_\nu \neq 0$ is nearly identical to the change to $\delta_{crit}$ using Eq.~(\ref{eq:deltacrit}). 

\bibliographystyle{ieeetr}
\bibliography{mdm}
\end{document}